\documentclass[10pt]{article}

\usepackage{amssymb,amsmath,amsfonts,eurosym,geometry,ulem,graphicx,caption,color,setspace,sectsty,comment,footmisc,pdflscape,subfigure,array,xurl,hyperref,authblk,placeins,booktabs,siunitx,threeparttable,adjustbox,longtable,multirow}

\usepackage[authoryear]{natbib}
\setcitestyle{round}

\usepackage[font=footnotesize]{caption}
\usepackage{caption}

\begin{document}

\begin{titlepage}
\title{\vspace{0.5cm}The congestion cost of electricity trade: Empirical\\evidence from German redispatch}
\author[a]{Alice Lixuan Xu\thanks{Corresponding author. E-mail: {alicelixuan@hotmail.com}}}
\affil[a]{\small\textit{Centre for Sustainability, Hertie School, Berlin}}
\date{\small Working Paper\\[0.5em]\today}

\maketitle
\begin{abstract}
This paper provides an empirical assessment of how European Union's policy for minimum trading capacities affects remedial congestion management within the German bidding zone. We study the implementation of the 70\% minimum remaining available margin (minRAM) requirement from mid-2022 to the end of 2025, using adjustments to ensure minRAM (AMR) as a measure of implementation intensity. Exploiting Germany’s structural North–South transmission bottleneck, we estimate the effect of AMR usage on redispatch volume using panel regressions, controlling for similar levels of system stress. We condition on trade via a data-driven post-double selection of neighbor-zone controls. We find that a 1 MW increase in AMR usage is associated with 0.2 MWh of additional hourly redispatch. Over the entire sample period, this corresponds to an estimated 10.8 TWh of additional redispatch and 1.9 billion euros in associated redispatch costs. The results provide empirical evidence for a grid–market gap in zonal electricity markets: when within-zone transmission constraints remain unresolved, increasing transmission capacity made available to the market increase the need for costly remedial congestion management outside the market.\\
\vspace{0in}\\
\noindent\textbf{Keywords:} Minimum trading capacity, congestion management, electricity market\\
\vspace{0in}\\
\noindent\textbf{JEL Codes:} C23, D47, Q48\\

\bigskip
\end{abstract}
\setcounter{page}{0}
\thispagestyle{empty}
\end{titlepage}
\pagebreak \newpage

\onehalfspacing

\section{Introduction} \label{sec:introduction}
Grid capacity allocation and congestion management are two sides of the same coin, central to electricity market design. In the interconnected zonal markets in Europe, trade and congestion between bidding zones on the day-ahead time frame are handled through flow-based market coupling (FBMC),\footnote{Flow-based market coupling was introduced in May 2015 to allocate trading capacities between the Belgian, Dutch, French, and German-Austrian bidding zones at the day-ahead market clearing stage, replacing the static net transfer capacity calculations to allow for more trade volume. Since June 2022, it has expanded to include eight more bidding zones to make up the Core Capacity Calculation Region that encompasses a large part of Central and Western Europe.} whose objective is to provide maximum available capacity to market participants within the security limits of the grid (\citeauthor{CommissionRegulationEU2015}, 2015/1222). At its core is an optimization problem characterized by the inherent tension between trade benefits and grid constraints. Meanwhile, congestion costs \textit{within} each bidding zone are, by design, not reflected in the zonal market outcome. Nevertheless, they are real costs for end consumers, who pay for remedial congestion management required to transform every infeasible market solution into a feasible dispatch, commonly via redispatch. For bidding zones that contain structural transmission bottlenecks, redispatch has become a routine and expensive process; in Germany, redispatch (including countertrading) currently costs 2.9 to 3.3 billion euros every year \citep{MonitoringReport2025}. Importantly, extensive redisaptch can create large differences between market results and actual physical flows, which can weaken the day-ahead market price signals' ability to inform locational scarcity, and further reduce socio-economic welfare \citep{poplavskayaIntegrationDayaheadMarket2020}.

As part of the European Union (EU)'s commitment to creating a fully integrated internal energy market, minimum remaining available margins (minRAMs) for electricity trading capacities are stipulated to increase and reach 70\% by the end of 2025 (\citeauthor{RegulationEU20192019}, 2019/943). In practice, when the minRAM criterion is not met, transmission system operators (TSOs) may adjust the network elements' remaining available margins (RAMs) to offer the market additional transmission capacity that is required by law but not physically there.\footnote{Besides short-run adjustments to the grid typology in FBMC, the other main way to fulfill the minRAM requirement would be the long-run grid expansion and upgrades \citep{schonheitMinimumTradingCapacities2021}. In this empirical study, we focus on the short-run adjustments actually made as a result of the minRAM requirement.} This artificial adjustment on trading capacities may have contributed to the hefty redispatch costs of recent years. Previous simulation studies projected that the additional redispatch costs, raised by the minRAM requirement, would outweigh the welfare gains it brings about \citep{bucksteegImprovingFlowbasedMarket2024,schonheitMinimumTradingCapacities2021}.

This paper is a first attempt at empirically quantifying the effect of minRAM on remedial congestion management within a bidding zone. We focus on the period from mid-2022 to the end of 2025 as the implementation of minRAM policy intensifies. We propose to measure minRAM implementation with the use of the artificial parameter AMR: Adjustment value to ensure Minimum Remaining available margins. We contribute to previous literature by identifying the increasing use of AMR as a driver of redispatch and estimate its effect using econometric and data-driven methods. While extensive load flow studies with detailed grid modeling provide a foundational understanding of the physical mechanism \citep{schonheitELMODDocumentationModeling2020, schonheitFundamentalUnderstandingFlowbased2021, feltenFlowbasedMarketCoupling2021, InsightsGermanysFuture, finckImpactFlowBased2021}, there are few empirical analyses of FBMC outcomes (e.g.,  \cite{ovaereEffectFlowbasedMarket2023}) or drivers of congestion within a bidding zone (e.g., \cite{titzIdentifyingDriversMitigators2024}) and none, to the best of our knowledge, on the empirical effect of the minRAM policy.

We find that on average, a 1 MW increase in AMR usage is empirically associated with a 0.2 MWh increase in hourly redispatch volume, given similar levels of system stress within the German bidding zone. A binary measurement of AMR activation strongly predicts the occurrence of redispatch, and when including both the binary and the continuous measurements of AMR to predict redispatch volume, the effect of AMR usage remains significant and at the same level.

This paper proceeds as follows. Section 2 provides the research background and the theoretical framework for this analysis, building on existing literature. Section 3 argues for the empirical strategy; Section 4 presents the data and measurement of the main variables; Section 5 expands on model variations and specifications. Section 6 presents our results, Section 7 discusses their limitations, and Section 8 concludes.

\section{Research background} \label{sec:background}
Based on theory and simulation studies, researchers agree that the minRAM policy increases trade benefits at the cost of increasing remedial congestion management \citep{bucksteegImprovingFlowbasedMarket2024, schonheitMinimumTradingCapacities2021, henneauxImpactAssessmentMinimum2021a, matthesImpactMinimumRemaining2019}. Specifically for the Core capacity calculation region, \cite{bucksteegImprovingFlowbasedMarket2024} project that minRAM increases cross-border trade by roughly 20 terawatt-hours but leads to a net welfare loss of 107 million euros annually due to additional redispatch. In many European zones, redispatch means curtailing renewable generation and ramping up conventional plants, as variable renewable systems are typically built in geographical locations away from demand centers \citep{schonheitMinimumTradingCapacities2021}.

The set of studies devoted to minRAM is rather limited, but a general insight we gain from the richer set of flow-based simulation studies is that the performance of the market strongly depends on how well it can represent the grid. \cite{kenisRegulatoryIncentivesTransmission2024a} emphasize that FBMC is prone to welfare losses because its performance depends on the parameters set by TSOs, and that the static and uniform minRAM design introduces inefficiency because the optimal RAM depends on spatially and temporally varying grid conditions that the market does not see. This point is echoed by \cite{weberImpactRegionalTimeDependent2023}, who propose a region- and time-dependent minRAM design that accounts for local grid constraints. Their simulation results suggest that a dynamic minRAM allows for a more cost-effective solution for market coupling: slightly curbing trade leads to significantly lower congestion management costs and a notable reduction in total system costs.

It is also important to mention active research on alternative market designs that can improve how grid constraints are considered in the market. This includes incorporating congestion management during market coupling \citep{poplavskayaIntegrationDayaheadMarket2020,schonheitFundamentalUnderstandingFlowbased2021, bucksteegImprovingFlowbasedMarket2024}, having accurate \citep{schonheitImpactDifferentStrategies2020} and dynamic generation shift keys (GSKs)\footnote{GSKs are node-level weights that distribute a marginal change in the zonal net position across the nodes within that zone. Or reversely, the contribution of an injection at a node to a change of the net position of the zone \citep{schonheitFundamentalUnderstandingFlowbased2021}. For readers familiar with FBMC inputs, multiplying nodal PTDF by a GSK matrix (derived from the D2CF base case) yields the zonal PTDF used in FBMC.} \citep{schonheitMinimumTradingCapacities2021}, creating grid-oriented capacity mechanisms \citep{InsightsGermanysFuture}, and improving zonal configurations \citep{fellingAssessingImprovedPrice2023,egererTwoPriceZones2016} or moving toward a nodal pricing system \citep{knorrZonalVsNodal2024}.

This paper aims to contribute to the broader conversation on electricity market design by quantifying a cost from widening the gap between how grid constraints are considered in the market and the grid reality, by examining the actual use of AMR in the implementation of increasing standards for minRAM. We take an econometric approach to estimate the causal effect of AMR on realized redispatch volume, asking:
\begin{itemize}
    \item On average, how did a greater AMR usage in day-ahead capacity calculations affect hourly redispatch volume within the four TSO control areas of the German bidding zone?
\end{itemize}

This question asks for empirical evidence that is lacking. Existing studies on the congestion impacts of FBMC are predominantly simulations, though with an increasing number of real-world details \citep{finckImpactFlowBased2021}. There are two main reasons for this. First, empirical data on remedial congestion management are rarely publicly available, standardized, or complete; extensive FBMC and the publication of its operational data are also rather recent phenomena \citep{jointallocationofficeJointAllocationOffice2025}. Second, any empirical method that seeks a causal explanation, within the established potential outcomes framework, is required to have cleanly defined treatment and control units that do not affect each other's treatment status (the so-called stable unit-treatment value assumption \citep{rubinRandomizationAnalysisExperimental1980}); this is not the case for nodes in a network. Therefore, a reduced-form empirical design that intentionally overlooks such grid structure is hardly preferred over the pursuit of a faithful representation of the grid, which can cleanly define counterfactuals that are modeled based on laws of physics. In order to study the messy real-world effects from minRAM, we propose a novel treatment variable based on the actual use of the AMR parameter in FBMC, allowing us to abstract away from the network structure to estimate the empirical effect on the aggregate level via static comparisons. Section \ref{sec:strategy} develops this empirical strategy.

We now recap the role of minRAM and AMR in relation to main parameters of FBMC, so as to understand the mechanism of how the use of AMR can weaken the correspondence between the market and the grid and lead to more congestion. We draw on existing literature, mainly \cite{schonheitMinimumTradingCapacities2021}. For more details on FBMC, see \citet{schonheitFundamentalUnderstandingFlowbased2021} for a brilliant introduction\footnote{For a shorter read, see also Section II of \cite{marjanovicImpactEnlargementFlow2018}.} and \cite{20250117CECCM} for relevant implementation rules.

On the electricity grid, physical flows do not follow commercial contracts, but rather laws of physics. Generation and consumption from various locations on the grid therefore affect and are constrained by individual elements' capacity on the entire network. The objective of FBMC is to map out a feasible domain that accounts for the constraints of all critical network elements (CNEs, e.g., transmission lines and substations) and find the optimal market outcome with the maximal welfare for the overall (Core) region. 

During FBMC, a change in the flow for each CNE is calculated based on two things: the changes in the net export positions (NP) of all coupled bidding zones and the CNE's sensitivity to each zone, i.e., its zonal power transfer distribution factors (PTDFs). The change in flow $f$ on CNE $c$ is a weighted sum of each coupled zone $z$'s change in its net position:
\begin{equation}
    f_{c}^{\Delta} = \sum_{z} NP_{z}^{\Delta} \cdot PTDF_{c, z}
\end{equation}
\noindent
where $\Delta$ denotes the change. Any change in flow for each CNE is bound by physical limitations of the grid. Therefore, TSOs compute RAMs to find the maximum transmission capacity that can be implicitly auctioned in the day-ahead market. A CNE's RAM is defined in both flow directions, by its maximum capacity $F_{max,c}$ (the maximum load it can carry), flow reliability margins $m_{c}$ (an uncertainty buffer for grid conditions and forecast errors), and reference flow $f_{ref,c}$ (the forecasted flow on the CNE). These parameters come from the common 2-Days Ahead Congestion Forecast (D2CF) base case, which reflects the likely loading of all network elements, given historical exchange on the reference day.
\begin{equation}
    RAM_{neg,c}\leq f_{c}^{\Delta} \leq RAM_{pos,c}
\end{equation}
\begin{equation}
    RAM_{pos,c} = F_{max,c} - m_c - f_{ref,c}
\end{equation}
\begin{equation}
    RAM_{neg,c} = -F_{max,c} + m_c - f_{ref,c}
\end{equation}
\noindent
The minimum limit on trading capacities, namely minRAM, comes in as the stipulated minimum RAM factor $\rho$ in between 20 and 70 \% multiplied by the CNE's maximum capacity; it sets the minimum value for RAM when there is no trade, i.e., all Core zones have zero net positions, such that
\begin{equation}
    \rho \cdot F_{max,c} = minRAM_{pos,c} \leq RAM_{pos,0,c} = F_{max,c} - m_c - f_{0,c}
\end{equation}
\begin{equation}
    \rho \cdot F_{max,c} = minRAM_{neg,c} \leq RAM_{neg,0,c} = -F_{max,c} + m_c - f_{0,c}
\end{equation}
where
\begin{equation}
    f_{0,c} = f_{ref,c} - \sum_{z} NP_{ref,z} \cdot PTDF_{c,z} 
\end{equation}
\noindent
The minRAM criterion is applied to RAM calculations under the zero-balance condition. Here, the so-called zero-balance flow $f_0$ represents what is left of $f_{ref}$ if taking out all associated flows from $NP_{ref}$, a counterfactual to the D2CF base case. Consequently, $f_0$ consists of possible flows from internal exchanges within the bidding zone and loop flows. Therefore, a 70\% minRAM rule means that those internal and loop flows, along with the flow reliability margins, should not take up more than 30\% of the CNE's maximum capacity, so as to leave at least 70\% of it available to the day-ahead market.

\begin{figure}[!p]
    \centering
    \includegraphics[width=1\textwidth,trim={3cm 3cm 3cm 2cm},clip]{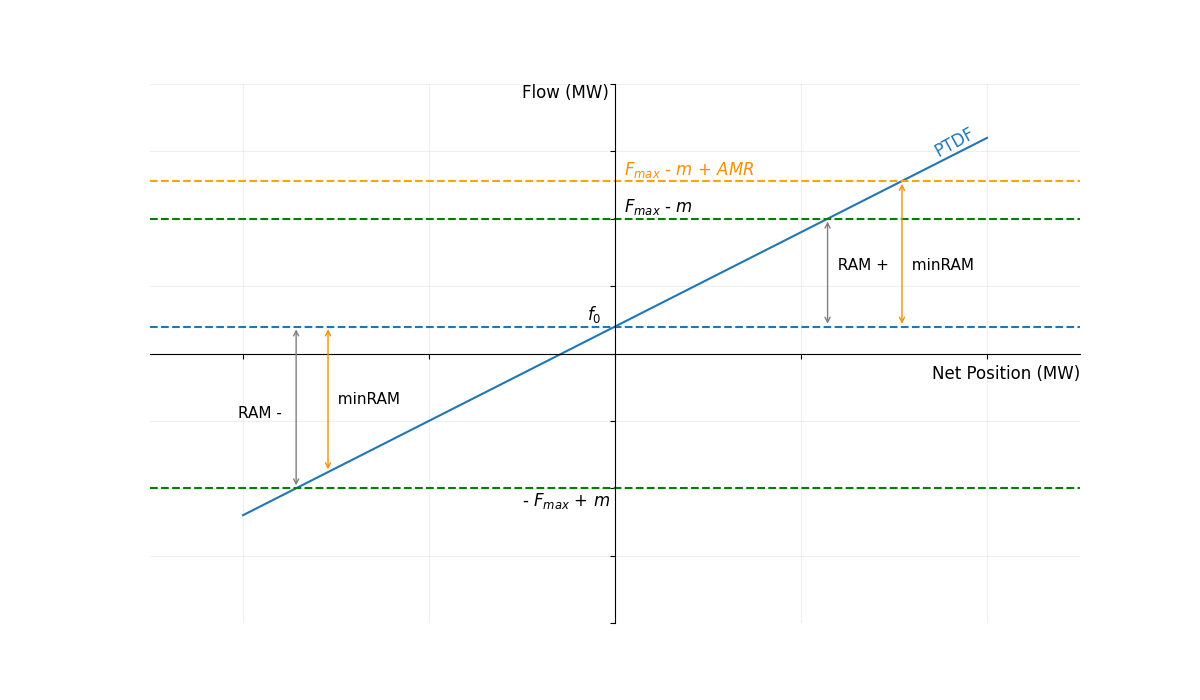}
    \caption{Illustration of a hypothetical CNE's RAM calculation and adjustment to ensure minRAM. The minRAM requirement is applied to RAM under zero-balance conditions, based on the CNE's maximum capacity $F_{max}$, flow reliability margins $m$, and zero-balance flow $f_0$. If such RAM, on either the positive or negative direction, is less than minRAM, say 70\% of $F_{max}$, it is then adjusted to minRAM by adding an adjustment value, AMR, on top. Such is the case for the positive RAM here; the negative RAM remains unadjusted because it is already larger than the same minRAM.}
    \label{fig:minRAM}
    \vspace{1em}
    \centering
    \includegraphics[width=1\textwidth,trim={0cm 0cm 0cm 0cm},clip]{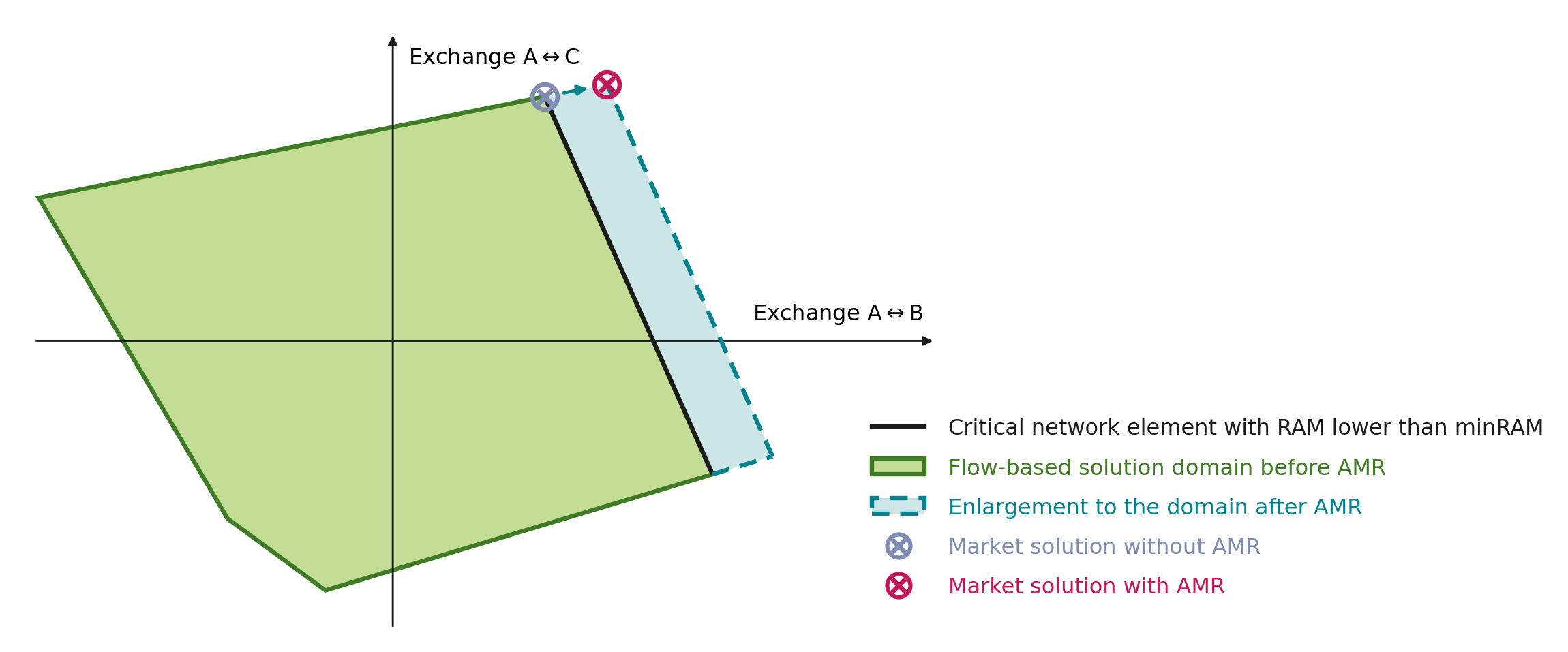}
    \caption{Stylized illustration of the effect of adding AMR to a domain-binding CNE within a three-zone scenario, where there are two interconnections, one between zones A and B and the other between zones A and C, following \cite{schonheitMinimumTradingCapacities2021}. Adding AMR to a domain-binding CNE enlarges the flow-based domain for market clearing and thus achieves its goal of granting more trading capacities to the day-ahead market. Here, we assume the added AMR is small enough that the adjusted CNE still binds the domain after the adjustment.} 
    \label{fig:amr_domain}
\end{figure}

Figure \ref{fig:minRAM} maps minRAM in relation to the aforementioned parameters for a hypothetical CNE. We see that when RAM in one direction is not large enough to fulfill the minRAM criterion, an adjustment value to ensure minRAM, i.e., AMR, is added on top to move the limiting bound in the CNE's RAM calculations, such that the adjusted RAM ensures the stipulated amount of transmission capacity.
\begin{equation}
    RAM_{pos,c} = F_{max,c} - m_c + AMR_{0,c} - f_{0,c}
\end{equation}

Figure \ref{fig:amr_domain} follows \cite{schonheitMinimumTradingCapacities2021} and illustrates the potentially congestion-inducing effect of AMR usage. Consider a three-zone scenario with two interconnections between zones A and B and between A and C, where we add a small AMR to a domain-binding CNE. As the added AMR artificially raises RAM without changing the CNE's physical capacity, it enlarges the flow-based domain but not the physical limit of the grid. Consequently, the cleared market solution in the enlarged domain may land in places that are physically infeasible, which then requires ex-post remedial congestion management to transform it to a feasible dispatch. This typically takes the form of redispatch: increasing generation in import-constrained nodes and curtailing generation in export-constrained nodes \citep{holmbergComparisonCongestionManagement2015, hirthMarketBasedRedispatchZonal2019}. To achieve redispatch, TSOs can instruct specific generators or groups of generators to adjust their output; they can also arrange offsetting cross-zonal exchange, i.e., countertrading, where the precise change to generation or load need not be pre-defined. In this study, we look at redispatch volume, including countertrading, to measure the need for remedial congestion management, or more commonly called in literature, redispatch needs.

During the implementation period from the start of Core capacity calculation in June 2022 to the end of 2025 where a strict 70\% minRAM requirement is stipulated to all CNEs, AMR is added based on TSO discretion, generally following the encouraged plan to have a linear increase in minRAM factors over time. We expect that this varying intensity in minRAM implementation over time is also reflected in the use of AMR, which would allow us to establish causal identification, the topic we turn to next.
\section{Empirical strategy} \label{sec:strategy}
We choose to study the German bidding zone for two main reasons. First, minRAM is potentially more congestion-inducing for the German bidding zone than for other European zones, due to its internal transmission bottlenecks. Remedial congestion management is less of a concern to begin with, for zones that configure transmission bottlenecks at the zonal borders and hence are able to price in structural congestion via the market, such as the Nordic ones. Second, Germany's persistent internal transmission bottlenecks allow for an empirical research design, in that the divide between the injection-heavy North and the consumption-heavy South allows for a reduced-form econometric model that can abstract away from the nodal structure and still preserve major congestion patterns. Being able to do that also helps tackle two main challenges for causal identification in this study: one is potential violation of the stable unit-treatment value assumption (SUTVA) and the other is mitigating omitted variable bias (OVB). We now discuss each in turn.

As mentioned before, causal inference under the established potential outcomes framework requires a clear identification of how treatment affects the outcome and simultaneously how the counterfactual non-treatment, ceteris paribus, would affect the outcome differently. This can be a challenge in a network of interconnected nodes because a change of flow on one CNE can change the flow on all other CNEs in the entire Core region; in a way, no unit can be considered fully untreated when any of them is. To account for such network interdependency, we define the observational unit at the TSO control-area level, not as individual nodes. By treating redispatch (and AMR) as an aggregate system outcome (and treatment), network spillovers within the unit of analysis should be greatly reduced, instead of constituting cross-unit interference. This specification thus reduces concerns about violating SUTVA. We investigate the potential bias from residual spillover effects on one control area from AMR used in the other three control areas in Section \ref{sec:result}.

Meanwhile, AMR may face OVB concerns because its application is up to TSO discretion, which may rely on information that also predicts realized redispatch needs. Consequently, unobserved drivers of congestion may affect both treatment assignment and outcomes. We therefore must identify and condition on relevant confounders, to mitigate OVB. Literature has established that in Germany, drivers of the North-South structural congestion include wind power generation in the North, electricity demand in the South, and cross-zonal trade \citep{titzIdentifyingDriversMitigators2024,egererTwoPriceZones2016,bucksteegImpactsRenewablesGeneration2015}. We hereby consider including those three types of drivers.

To control for wind power generation, actual output may be endogenous due to grid-related curtailment. Since the German Redispatch 2.0 reform in October 2021, solar and wind generators can be instructed to curtail production as part of redispatch measures. Observed generation output is thus downstream of redispatch orders, and instead, the upstream driver we need is rather generation potential. However, no publicly available data exist for this measurement, as available generation forecasts are based on historical generation output, which itself reflects historical curtailment. To circumvent this issue, we approximate wind generation potential by using zonal average wind speed scaled by control-area-specific installed capacity.\footnote{Due to data limitations, variation in weather and in capacity factors across wind parks of different locations are not considered.}

We include control-area grid load as another control variable. As demand is rather inelastic in Germany \citep{hirthHowAggregateElectricity2024}, we expect it to provide sufficient exogenous variation for the estimation of redispatch volume.

The story for trade, however, is more complex. Cross-zonal trade is a main outcome of the day-ahead market. Including it in any form (zonal net position, price, scheduled commercial exchange, etc.) risks including a mediator that absorbs the treatment effect in the model. We therefore aim to use contemporaneous neighbor-zone load and variable renewable generation, i.e., upstream drivers of trade, as a proxy for the exogenous variation in trade. Here, variable renewable technology types include solar photovoltaics, onshore and offshore wind power. Although all candidate variables from interconnected zones in the Core region may, in principle, affect Germany through complex physical and market interactions, we select and include a limited subset because many of them are highly correlated and additional controls could introduce too much multicollinearity to obtain stable estimates. This concern motivates a parsimonious baseline specification with German wind and load controls, which we then extend to our main specification by conditioning on trade, using a data-driven selection of neighbor-zone variables, as formalized in Section \ref{sec:method}.
\FloatBarrier
\section{Data and measurement} \label{sec:data}
We obtain hourly and quarter-hourly\footnote{Raw data from JAO publication tool and ENTSO-E transparency platform are of hourly granularity until the swtich to quarter-hourly on October 1, 2025.} FBMC final computation input data from JAO publication tool \citep{jointallocationofficeJointAllocationOffice2025}, monthly German control-area-level generation capacity and quarter-hourly load data from SMARD \citep{SMARDDownloadMarket}, hourly and quarter-hourly zonal load and technology-specific generation data from ENTSO-E transparency platform \citep{entso-eENTSOETransparencyPlatform2025}, hourly weather data from NASA MERRA-2 \citep{MERRA2Tavg1_2d_flx_Nx2d1HourlyTimeAveragedSingleLevelAssimilationSurface} and German redispatch event data from Netztransparenz \citep{NetztransparenzSystemdienstleistungenBetriebsfuehrung}. We transform and harmonize all data into an hourly time-series format for the sample period from June 9, 2022 to December 31, 2025. We include a complete table of summary statistics in the Appendix. In the rest of this section, we discuss measurement for the outcome and treatment variables.

To measure redispatch, we use current-related redispatch and countertrading orders in aggregate MWh volume. We drop order requests that were issued for tests and voltage-related reasons because they are independent from congestion; this step is consistent with \cite{titzIdentifyingDriversMitigators2024}. We also drop order requests that did not come from any of the four German TSOs, meaning that the congestion they aimed to resolve was not in any of the four control areas, following the definitions given by \cite{BundesnetzagenturBeschlusskammer6Festlegungsverfahren}. We then transform the order data into a time-series format, summing up redispatch volume that correspond to each hour, in both up- and downward directions. We also explore model variations on the direction-specific decomposition of the outcome and include their results in the Appendix.

\begin{figure}
    \centering
    \includegraphics[width=1\textwidth,trim={0cm 0cm 0cm 0cm},clip]{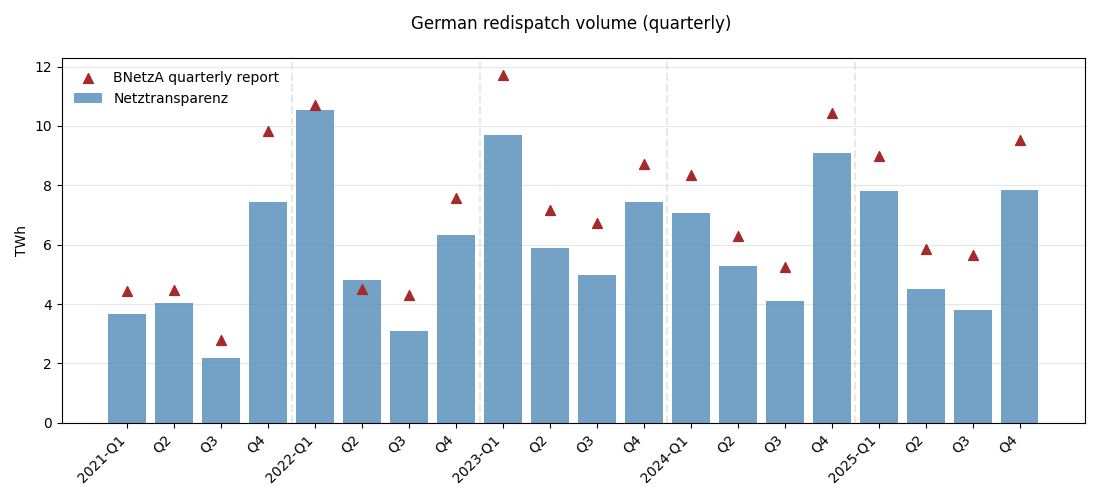}
    \caption{Detailed redispatch order data from Netztransparenz are aggregated to quarterly sums, including both up- and downward redispatch, and countertrading, for a like-for-like comparison against the official numbers published by Bundesnetzagentur (BNetzA) in its quarterly monitoring reports.} 
    \label{fig:rdp_quarterly}
\end{figure}

The data quality on German redispatch has improved in recent years but is not perfect. Germany's national regulatory aganecy, Bundesnetzagentur, publishes quarterly reports on redispatch and switched to the central SMARD platform in 2024 \citep{BundesnetzagenturNetzengpassmanagement, SMARDNetzengpassmanagement}. Figure \ref{fig:rdp_quarterly} compares the data we use for this study against the quarterly numbers published by Bundesnetzagentur; we see that the two agree on the general trend but the quarterly sums from Netztransparenz tend to fall short and therefore could be incomplete. This limitation from data quality potentially poses a downward bias on our model results.

We also find that the zonal aggregate up- and down-ward redispatch volumes differ by more than 50 MW for 40\% to 47\% of the hours in each year of the sample. This asymmetry may arise from cross-border redispatch, though FBMC should account for (most of) cross-border congestion.\footnote{Cross-border congestion can arise from forecast errors of the 2DCF base case; it can also arise from the use of AMR on cross-border CNEs, which we do not include in this analysis.} Alternatively, redispatch may have been used to balance market demand and supply, despite not being designed for balancing purposes. As the contemporaneous Pearson correlations of the imbalance price with redispatch volume and AMR usage are small, respectively at -0.06 and -0.08, we consider it unlikely to threaten our empirical strategy and thus do not investigate this asymmetry further in this paper.

To measure AMR usage, we take the sum of base-case AMR (in MW) applied to all CNEs in each control area in each hour. We take the absolute value of AMR, not relative to each CNE's capacity, because AMR is artificial MW added on top of the CNE's RAM to fill its distance to minRAM, independent of the CNE's maximum capacity. This thus captures the aggregate intensity of AMR intervention on the input constraints of the most-likely case used for day-ahead market clearing.

Both redispatch and AMR data have a substantial share of zero observations, as neither process is expected to occur in every hour, particularly when grid load is low and congestion unlikely. Table \ref{tab:stats_rdp_amr} presents summary statistics for both variables. We see that a large share of zero observations is common across all German TSO control areas. By controlling for load and upstream drivers of congestion, we account for expected zero observations; we address additional zero-inflation concerns through robustness checks. 
\begin{table}[htbp]
\centering
\caption{Summary statistics for hourly aggregate redispatch\\and AMR usage by TSO control area}
\label{tab:stats_rdp_amr}
\small
\begin{tabular}{lrrrrr}
\toprule
 & \% of zero & Mean & 75\% & Max & Std. \\
\midrule

\textbf{Redispatch volume (MWh)} & & & & & \\
\quad 50Hertz    & 54.3 & 386 & 425   & 7,857  & 762 \\
\quad Amprion    & 49.1 & 603 & 873   & 6,976  & 962 \\
\quad TenneT DE  & 40.8 & 882 & 1,190 & 11,991 & 1,412 \\
\quad TransnetBW & 47.5 & 365 & 546   & 3,905  & 571 \\
\hline
\quad \textit{All control areas (pooled)} & 47.9 & 559 & 703 & 11,991 & 1,000 \\
\hline
\addlinespace
\textbf{AMR usage (MW)} & & & & & \\
\quad 50Hertz    & 31.3 & 742 & 998 & 16,290 & 1,292 \\
\quad Amprion    & 53.9 & 289 & 241 & 8,787  & 683 \\
\quad TenneT DE  & 35.2 & 645 & 877 & 9,551  & 1,055 \\
\quad TransnetBW & 68.4 & 68  & 51  & 2,253  & 157 \\
\hline
\quad \textit{All control areas (pooled)} & 47.2 & 436 & 414 & 16,290 & 944 \\

\bottomrule
\end{tabular}
\end{table}

\section{Model specifications} \label{sec:method}
We start with the parsimonious baseline: a panel ordinary least squares (OLS) regression that exploits the variation in AMR usage over time within TSO control areas, controlling for TSO and hour-of-day fixed effects and minimal controls. For the main model specification, we additionally condition on trade by including a data-driven selection of control variables from other bidding zones in the Core region.
\subsection{Parsimonious baseline}
To estimate the empirical association between AMR usage and redispatch volume within the German bidding zone, we begin with a parsimonious panel OLS specification, conditioning on two main drivers of system stress: load and wind generation potential. Let \(i\) index the four TSO zones and \(t\) index hours. The baseline model is thus
\begin{equation}
\label{eq:baseline}
Y_{it}=
\beta_0 +
\beta_1 D_{it}
+
\beta_2 L_{it}
+
\beta_3 W_{it}
+
\alpha_i
+
\gamma_{h(t)}
+
\varepsilon_{it}
\end{equation}
\noindent
where in TSO control area \(i\) at hour \(t\), \(Y_{it}\) denotes the aggregate up- and downward redispatch volume (in MWh), \(D_{it}\) denotes treatment, i.e., the total AMR applied (in MW), \(L_{it}\) denotes load (in MW), and \(W_{it}\) denotes wind generation potential and is approximated by wind speed scaled by installed wind capacity (MW$\cdot$m/s). Here, wind generation potential for onshore and offshore wind energy are separate variables; this distinction is not shown in the equation for the sake of simplicity. The model includes entity fixed effects for each TSO control area within the bidding zone, \(\alpha_i\), to control for time-invariant differences in network characteristics across TSO control areas and hour-of-day fixed effects, \(\gamma_{h(t)}\), which absorb diurnal patterns\footnote{We also explored model variations that include day-of-week, month, and year dummy variables, and find that only the hour-of-day fixed effects are significant.} common to all control areas. \(\varepsilon_i\) is the residual error term.
\subsection{Main model with post-double selection}
To select additional control variables from across the interconnected Core region, we use post-double selection (PDS), as established by \cite{belloniHighDimensionalMethodsInference2014}. The method is designed to select a sparse set of controls that protect inference against omitted variables that may be relevant drivers of either the outcome or the treatment.

Let $Z_t$ denote the vector of candidate neighbor-zone controls, and $D_{it}$ denotes the treatment and $Y_{it}$ the outcome. The candidate controls are indexed by $t$ as they vary over time but are common across German TSO control areas within a given hour. Prior to selection, the outcome, treatment, and candidate controls are residualized with respect to the baseline controls and the fixed effects. This step is to make sure that additional variables are selected for their explanatory power in addition to that of the baseline controls and the fixed effects. The PDS procedure then applies LASSO\footnote{Least absolute shrinkage and selection operator, formally introduced by Equation (12).} separately to the residualized outcome and treatment equations:
\begin{align}
\widetilde{Y}_{it}
    &= \widetilde{Z}_{it}'\pi_Y + u_{it} \\
\widetilde{D}_{it}
    &= \widetilde{Z}_{it}'\pi_D + v_{it}
\end{align}
\noindent
Here, tildes denote residuals after partialling out the baseline controls and fixed effects. Although the candidate variables $Z_t$ are common across TSOs within each hour, the residualized variables are denoted by $\widetilde{Z}_{it}$ because partialling out TSO-specific baseline controls may induce TSO-specific residual variation.

For each of the two selection equations (10) for outcome $Y$ and (11) for treatment $D$, LASSO estimates the coefficient vector by solving
\begin{equation}
\widehat{\pi}(\lambda)
=
\arg\min_{\pi}
\left\{
\frac{1}{N}\sum_{i,t}
\left(
\widetilde{Q}_{it}
-
\widetilde{Z}_{it}'\pi
\right)^2
+
\lambda \sum_{j=1}^K |\pi_j|
\right\}
\end{equation}
\noindent
where $\widetilde{Q}_{it}$ denotes either the residualized outcome $\widetilde{Y}_{it}$ or residualized treatment $\widetilde{D}_{it}$, and $\widetilde{Z}_{t}$ remains the vector of residualized candidate controls. Prior to LASSO estimation, the residualized outcome, treatment, and candidate controls are standardized to zero mean and unit variance. $N$ denotes the total number of observations, $j$ indexes the individual candidate controls, and $K$ is the total number of candidate controls. The vector \(\pi=(\pi_1,\ldots,\pi_K)'\) contains the coefficients on the candidate controls over which the LASSO objective is minimized, while \(\widehat{\pi}(\lambda)\) denotes the resulting estimated coefficient vector for a given penalty parameter \(\lambda\). Its \(j\)-th element,\(\pi_j\), is the coefficient associated with candidate control \(j\). The penalty \(\lambda\sum_{j=1}^K|\pi_j|\) creates a trade-off between model fit and sparsity:  assigning a non-zero coefficient to a candidate control may reduce the residual sum of squares, but it also increases the penalty term. Whenever the improvement in model fit is not large enough to offset the penalty, LASSO finds the optimal coefficient at zero, thereby excluding the corresponding candidate variable from the selected set.

The penalty parameter $\lambda$ hence determines the degree of regularization. We consider two approaches for deciding this regularization parameter, using five-fold cross-validation, where observations were divided into five consecutive time blocks (as opposed to randomly assigned to folds). The \textit{CV-min} rule selects the value of $\lambda$ that yields the lowest mean prediction error across the five validation folds. Alternatively, the \textit{one-standard-error (1-SE)} rule allows the mean cross-validation error to increase by up to one standard error above the previous minimum and among the values that satisfy this criterion, selects the largest $\lambda$. Because a larger $\lambda$ imposes stronger regularization, the 1-SE rule typically selects a sparser set of controls while retaining prediction performance close to the cross-validation optimum.

Let $\widehat{S}_Y$ and $\widehat{S}_D$ denote the index sets selected in the outcome and treatment equations, respectively. The final PDS conditioning set is their union:
\begin{equation}
\widehat{S}
=
\widehat{S}_Y \cup \widehat{S}_D
\end{equation}
\noindent
which then enters the unpenalized fixed-effects regression:
\begin{equation}
\label{eq:pds}
Y_{it}
=
\beta_0
+
\beta_1 D_{it}
+
\beta_2 L_{it}
+
\beta_3 W_{it}
+
Z_{t,\widehat{S}}'\delta
+
\alpha_i
+
\gamma_{h(t)}
+
\varepsilon_{it}
\end{equation}
\noindent
where $\delta$ is the vector of coefficients on the PDS-selected neighbor-zone controls. Hats indicate that the selected sets are decided based on the data, as opposed to the ground truth. The union step is important because a candidate variable may only weakly predict the treatment but strongly predict the outcome, such that selecting controls solely from the treatment equation may omit relevant confounders \citep{belloniHighDimensionalMethodsInference2014}. Since candidate neighbor-zone controls are common across TSO control areas, they cannot explain contemporaneous cross-TSO variation in AMR, but they may still explain its temporal variation.

\section{Results} \label{sec:result}
\begin{figure}
    \centering
    \includegraphics[width=0.8\textwidth,trim={0cm 0cm 0cm 0cm},clip]{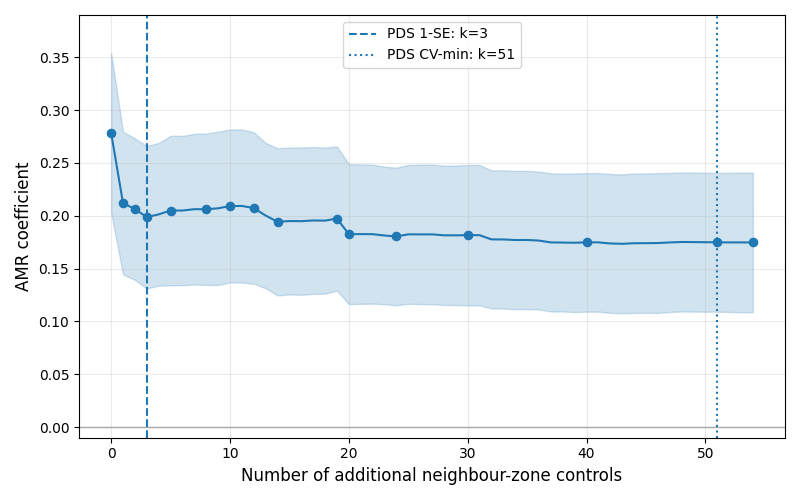}
    \caption{Estimated AMR coefficients as neighbor-zone controls are added (in the order of their entry) to the baseline specification along the PDS LASSO path. All specifications include German control-area grid load, onshore and offshore wind generation potential, TSO and hour-of-day fixed effects are included. Here, $k$ denotes the number of neighbor-zone controls included. The shaded area shows 95\% confidence intervals. The dashed line marks the control set selected by the one-standard-error rule ($k=3$), and the dotted line marks the set selected by minimum cross-validation error ($k=51$), both based on five-fold time-blocked cross-validation.} 
    \label{fig:pds}
\end{figure}

Across all model specifications, we find a positive and statistically significant association between AMR usage and redispatch volume within the German bidding zone. Table \ref{tab:results} shows the parsimonious baseline model result on the left, and that of the PDS adjusted specification with neighbor-zone controls on the right. Three additional controls on onshore wind generation from Denmark and Poland are selected from the PDS procedure under (1-SE) regularization. The effect of AMR attenuates but stays positive and significant after including additional controls. Figure \ref{fig:pds} also gives a view on the sensitivity of model results to including more neighbor-zone controls. We see that after including the selected three additional controls, the effect of AMR usage stabilizes.
\begin{table}[htbp]
    \centering
    \caption{Effect of AMR on aggregate German redispatch}
    \label{tab:results}
    \small
    \begin{tabular}{lcc}
        \toprule
        & (1) Parsimonious baseline
        & (2) PDS (1-SE) adjusted \\
        \midrule

        AMR
        & $0.279^{***}$
        & $0.199^{***}$ \\
        & $(0.039)$
        & $(0.034)$ \\[0.3em]

        AMR effect per IQR (414 MW)
        & $115.3$ MW
        & $82.3$ MW \\

        \midrule
        Additional PDS controls
        & $0$ & $3$ \\

        TSO fixed effects
        & Yes & Yes \\

        Hour-of-day fixed effects
        & Yes & Yes \\

        Observations
        & $123{,}689$
        & $123{,}689$ \\

        Within $R^2$
        & $0.272$
        & $0.408$ \\

        \bottomrule
    \end{tabular}

    \vspace{0.5em}
    \begin{minipage}{0.90\textwidth}
        \footnotesize
        \textit{Notes:} The dependent variable is hourly redispatch volume in MWh. Column (1) reports the parsimonious baseline specification, controlling for TSO-level grid load and onshore and offshore wind generation potential. Column (2) shows the main model specification which augments the parsimonious baseline model with three neighbor-zone controls selected by the one-standard-error post-double-selection (PDS) LASSO rule: onshore wind generation in DK1, DK2, and in Poland. Driscoll-Kraay standard errors are reported in parentheses and computed using a Bartlett kernel with bandwidth 336. The coefficients on the intercept, fixed effects, and additional PDS controls are omitted for brevity.
        $^{*}\,p<0.05$,
        $^{**}\,p<0.01$,
        $^{***}\,p<0.001$.
    \end{minipage}
\end{table}

In the main specification (column (2), Table \ref{tab:results}), a 1 MW increase in AMR usage is found to be associated with a 0.20 MWh increase in the hourly up- and downward redispatch volume. Moving from the 25th (0 MW) to the 75th percentile (414 MW) of AMR usage corresponds to about 83 MWh higher expected redispatch within the hour, equivalent to about 15\% of average hourly redispatch volume in the sample. The estimate is identified from within-TSO variation over time, conditional on control-area load, onshore and offshore wind generation potential, TSO and hour-of-day fixed effects, as well as onshore wind generation from Danish and Polish zones.
\FloatBarrier
\subsection{Robustness}
The positive association between AMR usage and redispatch volume is robust when accounting for the substantial mass at zero in both variables. Table \ref{tab:robustness} compares main results to those of 4 additional model specifications. We first look at the binary and continuous measurements of AMR usage, before turning to those of redispatch volume. Adding an indicator for positive AMR usage (AMR activation) leaves the estimated treatment (AMR intensity) coefficient virtually unchanged from the main results (specification [1]). Furthermore, restricting the sample to hours with positive AMR yields a similarly sized coefficient at 0.17 (specification [2]). This suggests that the baseline relationship is not driven solely by the differences between hours with and without AMR activation.

Separating redispatch occurrence from its magnitude informs us that AMR usage affects the two aspects differently. We find that AMR activation is associated with an approximately 9.2 percentage-point higher probability of positive redispatch, whereas the amount of AMR usage is not significantly associated with the probability of redispatch once activation is accounted for (specification [3]). When conditioning on positive redispatch, however, AMR intensity remains positively and statistically significantly associated with redispatch volume, with an estimated coefficient of 0.17 (specification [4]). These results suggest that binary AMR activation can explain the occurrence of redispatch, and that the continuous intensity of AMR usage provides additional explanation for the magnitude of redispatch volume.
\begin{table}[!htbp]
\centering
\caption{Robustness to alternative model specifications}
\label{tab:robustness}

\small

\begin{tabular}{p{7.0cm}ccc}
\toprule
Model specification
& AMR activation
& AMR intensity
& Observations \\
\midrule

\textbf{Main}
\\full sample\\Dependent variable: redispatch volume
& --
& $0.199^{***}$
& 123,689 \\
& --
& $(0.034)$
& \\[0.5em]

\textbf{[1] AMR activation vs. intensity:}
\\full sample\\Dependent variable: redispatch volume
& $51.649^{**}$
& $0.192^{***}$
& 123,689 \\
& $(25.336)$
& $(0.035)$
& \\[0.5em]

\textbf{[2] AMR intensity:}
\\subsample when $AMR_{it}>0$\\Dependent variable: redispatch volume
& --
& $0.170^{***}$
& 65,324 \\
& --
& $(0.035)$
& \\[0.5em]

\textbf{[3] Redispatch occurrence:}
\\full sample, linear probability model\\Dependent variable: $\mathbf{1}(Y_{it}>0)$
& $0.092^{***}$
& $0.000008$
& 123,689 \\
& $(0.014)$
& $(0.000009)$
& \\[0.5em]

\textbf{[4] Redispatch magnitude:}
\\subsample when $Y_{it}>0$\\Dependent variable: redispatch volume 
& $65.356^{*}$
& $0.166^{***}$
& 64,412 \\
& $(37.985)$
& $(0.039)$
& \\[0.5em]


\bottomrule
\end{tabular}

\vspace{0.5em}

\begin{minipage}{0.97\textwidth}
\footnotesize
\textit{Notes:}
Standard errors are reported in parentheses. AMR activation is defined as $\mathbf{1}(D_{it}>0)$.
AMR intensity refers to continuous AMR usage measured in MW. All specifications include the baseline and neighbor-zone controls selected in the main specification, and TSO and hour-of-day fixed effects. All models are panel OLS except for one panel linear probability model with binary redispatch outcome.\\$^{***}p<0.01$, $^{**}p<0.05$, $^{*}p<0.1$.
\end{minipage}

\end{table}
\begin{table}[htbp]
\centering
\caption{Regression results for subsamples by year}
\label{tab:amr_yearly}
\small
\begin{tabular}{lrrrrrr}
\toprule
Year & AMR coefficient & Std. Error & Observations & $R^2$ \\
\midrule
2022 & 0.372 & 0.264 & 19,469 & 0.354 \\
2023 & 0.280*** & 0.079 & 34,664 & 0.447 \\
2024 & 0.180*** & 0.048 & 34,784 & 0.409 \\
2025 & 0.289*** & 0.040 & 34,772 & 0.432 \\
\bottomrule
\end{tabular}

\vspace{0.2cm}
\begin{minipage}{0.92\textwidth}
\footnotesize
\textit{Notes:} The dependent variable is aggregate hourly redispatch in MWh. All specifications include TSO and hour-of-day fixed effects, and the same set of control variables as the main specification. The 2022 sample include observations from June 9th onward. *** $p<0.01$, ** $p<0.05$, * $p<0.10$.

\end{minipage}
\end{table}

Additionally, to check if the main result is driven by one particular year, we run the main specification model separately for each year in the sample period and present the regression results in Table \ref{tab:amr_yearly}. We see that aside from year 2022, a positive and significant association between AMR usage and redispatch volume is present throughout the sample period. The estimated result from year 2022 is also positive, but not statistically significant. This may reflect its less variation and volume in AMR usage as the required minRAM factors were at its lowest (at 31\%), compared to the following years. The 2022 subsample is also distinctive in that it coincides with the peak of the European energy crisis, where drastic changes in power plant dispatch and unexpected failures in neighbor-zone generation could have increased noise in redispatch, unrelated to minRAM implementation.
\subsection{Control-area spillover}
We cannot assume that higher AMR usage in one TSO control area does not simultaneously alter the grid conditions for its neighbor control areas. Therefore, we augment the main model with contemporaneous AMR usage in the other German TSO control areas. First, we include four TSO-specific AMR variables, setting the one corresponding to the observation's own TSO to zero. Second, we replace the individual other-TSO variables with the sum of AMR usage in all other German TSO control areas.

The main AMR coefficient drops from 0.20 to 0.18 when the aggregate other-TSO AMR control is included, while remaining highly statistically significant (column (3), Table \ref{tab:robustness_other_amr}). We think this drop in AMR coefficient may reflect the indirect pathway where one TSO's AMR usage induces its neighbors' AMR usage, which then affect redispatch in its own area. Additionally, we find that none of the individual other-TSO AMR coefficients is significant (column (2), Table \ref{tab:robustness_other_amr}). This pattern suggests that the AMR effect operates in a system-wide fashion, more so than through distinct pairwise spillovers between specific TSO control areas.
\begin{table}[!htbp]
\centering
\caption{Effect of contemporaneous AMR usage in other TSO control areas}
\label{tab:robustness_other_amr}
\small
\begin{tabular}{p{5.2cm}ccc}
\hline
&
\begin{tabular}{c}
Main
\end{tabular}
&
\begin{tabular}{c}
Individual other-TSO \\
AMR controls
\end{tabular}
&
\begin{tabular}{c}
Sum of other-TSO \\
AMR control
\end{tabular}
\\
\hline

AMR
& $0.199^{***}$
& $0.157^{***}$
& $0.178^{***}$ \\
& $(0.034)$
& $(0.033)$
& $(0.028)$ \\[0.4em]

Other AMR: 50Hertz
& --
& $0.007$
& -- \\
& --
& $(0.019)$
& -- \\[0.4em]

Other AMR: Amprion
& --
& $0.018$
& -- \\
& --
& $(0.037)$
& -- \\[0.4em]

Other AMR: TenneT DE
& --
& $0.024$
& -- \\
& --
& $(0.029)$
& -- \\[0.4em]

Other AMR: TransnetBW
& --
& $0.394$
& -- \\
& --
& $(0.243)$
& -- \\[0.4em]

Other AMR: Sum
& --
& --
& $0.022^{**}$ \\
& --
& --
& $(0.011)$ \\

\hline

Observations
& 123,689
& 123,689
& 123,689 \\

Within $R^2$
& 0.408
& 0.411
& 0.409 \\

\hline
\end{tabular}

\vspace{0.5em}

\begin{minipage}{0.96\textwidth}
\footnotesize
\textit{Notes:}
The dependent variable is aggregate hourly redispatch volume in MWh. Standard errors are reported in parentheses. Column (2) additionally includes four TSO-specific variables capturing contemporaneous AMR usage in the German TSO control areas; for each observation, the AMR variable corresponding to the observation's own TSO is set to zero. Column (3) instead controls for the sum of contemporaneous AMR usage across the other TSO control areas. All specifications include control-area load, onshore and offshore wind generation potential, TSO and hour-of-day fixed effects, as well as the neighbor-zone controls selected by PDS under the 1-SE LASSO specification.\\$^{***}p<0.01$, $^{**}p<0.05$, $^{*}p<0.1$.
\end{minipage}
\end{table}
\subsection{Cost implications}
To estimate the redispatch costs associated with AMR usage, we multiply the associated redispatch volume estimate by an average cost of redispatch in Germany.

\begin{figure}
    \centering
    \includegraphics[width=1\textwidth,trim={0cm 0cm 0cm 0cm},clip]{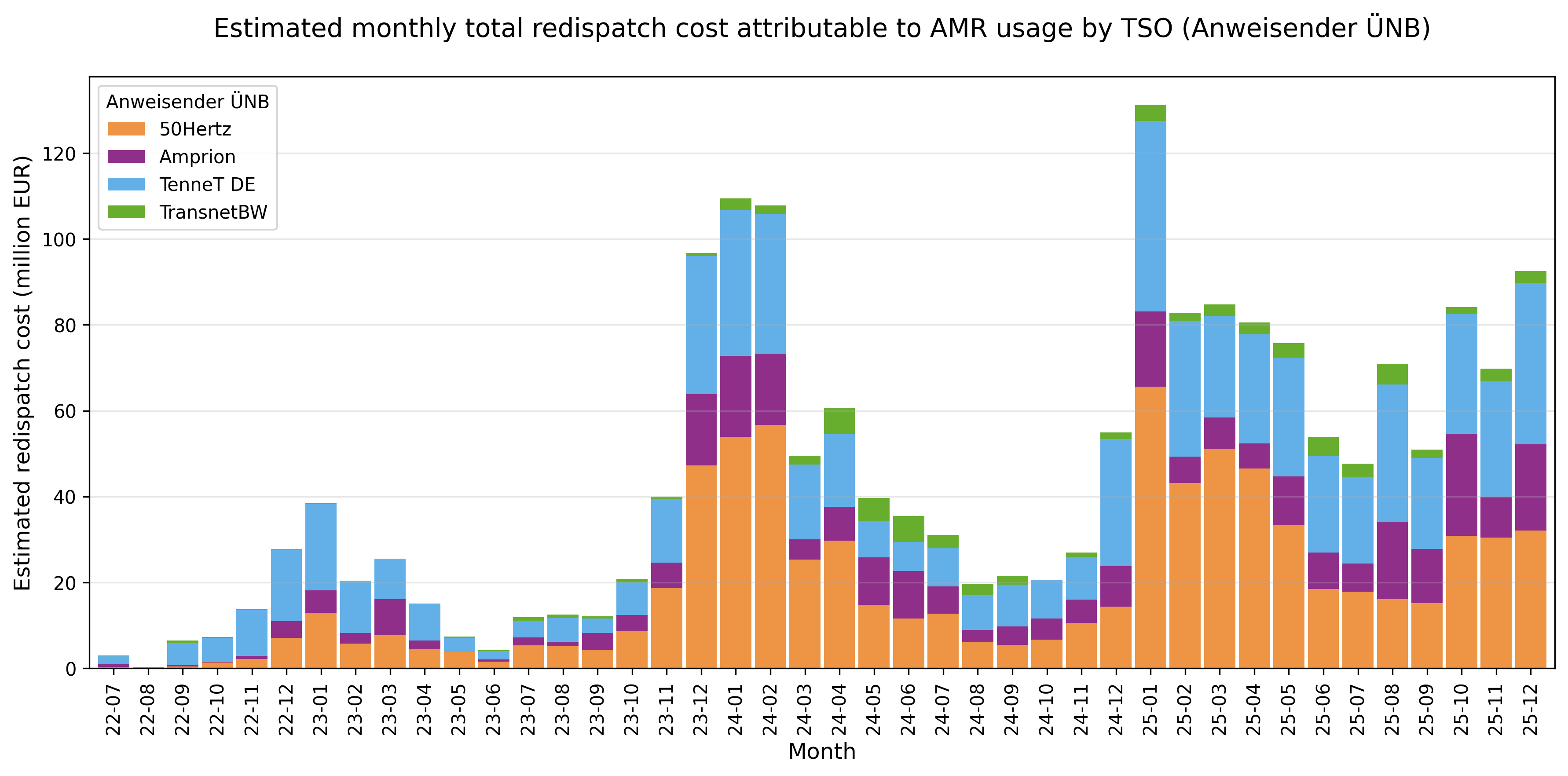}
    \caption{Estimated redispatch cost that is attributable to AMR usage, monthly total. The data are color-coded by the TSO that implements the redispatch within its control area. Public data on the unit cost of redispatch starts from July 2022 \citep{SMARDRedispatchKostena}.} 
    \label{fig:estimated_cost}
\end{figure}

\cite{SMARDRedispatchKostena} publishes monthly average redispatch costs per MWh by two general technology types: conventional and renewable. We thus map redispatch volume data to the two technology types and calculate the average cost of redispatch as a monthly volume-weighted average. Redispatch cost that can be attributed to AMR usage is therefore
\[
\widehat{C}_{i,m}
=
\overline{c}_{m}\cdot \beta_{1}
\sum_{t \in m} D_{i,t}
\]
\noindent
where $\mathrm{D}_{i,t}$ denotes MW AMR usage by control area $i$ at hour $t$, as before; $\beta_1$ denotes the estimated AMR coefficient from main results, at 0.199; $\overline{c}_{m}$ expresses the weighted average cost of redispatch per MWh, constant within each month $m$. Figure \ref{fig:estimated_cost} shows monthly total costs attributable to AMR usage by TSO. With the steady increase in AMR usage (see Figure \ref{fig:amr} in Appendix), the annual total associated redispatch cost is estimated to have climbed from 305 in 2023, 577 in 2024, to 925 million euros in 2025.
\FloatBarrier
\section{Discussion} \label{sec:discussion}
Maximizing trade between interconnected zonal markets is undoubtedly an essential objective of European electricity market design. We do not seek to downplay that, nor the welfare gains that the minRAM policy may have achieved to that end. Rather, this paper is limited to quantifying the cost side: the additional outside-the-market redispatch required in Germany to implement the minRAM policy. By doing so, we highlight the persistent issue of transmission bottlenecks and show that congestion within the German bidding zone can be exacerbated by policies designed to promote cross-border trade.

In principle, Member States are expected to resolve internal transmission bottlenecks by developing their grid or by improving their zonal configurations to better reflect physical constraints during market coupling. Germany has largely maintained the existing zonal structure while artificially augmenting its trading capacities made available to the market on an ad-hoc basis. This widens the gap between the market representation of the German grid and its physical reality, making its end consumers bear the additional costs of remedial congestion management.

With this analysis, we cannot determine to what extent the additional redispatch associated with AMR usage was due to transmission bottlenecks within the German bidding zone, as we do not observe the counterfactual: a German bidding zone without its transmission bottlenecks. Future research could address this limitation through a comparative analysis across multiple bidding zones, whose transmission constraints are more or less accurately represented in the market-coupling process. This would allow researchers to examine whether internal transmission bottlenecks moderate the effect of AMR usage on redispatch. For instance, we expect a bidding zone that places its major transmission bottlenecks at its borders to be able to price in most of its congestion during zonal market coupling. Consequently, there will be less need for redispatch and AMR usage may also have a weaker effect in such bidding zones.

This analysis is also limited to AMR usage on CNEs within the German bidding zone, not including cross-border CNEs. For within-zone CNEs, AMR usage reflects the German minRAM implementation and the relief of any resulting congestion is coordinated by German TSOs and recorded in German redispatch data. For cross-border CNEs, meanwhile, capacity limits can be set jointly by the two TSOs sharing the network element, hence reflecting a common minRAM implementation plan, and the relief of its congestion may also be coordinated outside of German redispatch. Including cross-border CNEs may therefore introduce mismeasurement in the treatment and partial non-observation of the outcome, both of which would attenuate the estimate toward zero.

\section{Conclusion} \label{sec:conclusion}
Congestion between bidding zones is explicitly accounted for during day-ahead market clearing, whereas congestion within bidding zones is largely left to be resolved after market clearing through remedial congestion management. This creates an inherent tension between cross-border trade and local congestion. Increasing trading capacities can improve market integration and price convergence, at the cost of aggravating local congestion within affected bidding zones. 

In Germany, redispatch costs are already high as its zonal configuration in the market masks its structural transmission bottlenecks on the grid. Because the costs of remedial congestion management arise outside the day-ahead market, the increasing use of ad-hoc adjustment parameters such as AMR can further weaken the correspondence between the market representation of the grid and the underlying physical constraints during FBMC. In this way, improvements in market efficiency hardly translate to more efficient local system operation; instead, they are achieved at the cost of greater and more expensive remedial intervention within the local bidding zone. Reliance on such short-term solutions therefore risks widening the gap between the market and the grid, undermining the central objective of electricity market design: that market outcomes should reflect the location and scarcity of network capacity and ultimately, the economic value of electricity.

This paper provides an empirical quantification of the effect of AMR usage on redispatch. We find that in Germany, each additional 1 MW CNE capacity artificially augmented to ensure minRAM raises, on average, 0.2 MWh additional redispatch, across TSO control areas, as the minRAM requirement tightened between mid-2022 and the end of 2025. This corresponds to 10.8 TWh of total additional redispatch over the 3.5-year sample period, and about 1.9 billion euros.

The tension between trade and congestion becomes increasingly relevant as the European power system decarbonizes. While grid expansion is slow, the rapid deployment of renewable generation distant from demand centers is increasing both redispatch needs and renewable curtailment. Policies that stipulate cross-zonal trading requirements without changing the underlying internal network constraints may therefore exacerbate both.

More broadly, our findings also highlight a distributional feature of the European market-integration framework. The minRAM requirement results in congestion-management costs that are largely borne within the affected bidding zones rather than reflected in the cross-zonal market outcome that generates them. This creates pressure on highly congested bidding zones either to expand their internal grids, reconsider bidding-zone configurations, or continue to absorb increasing redispatch costs.

\subsection*{Data availability}
The code and data needed to replicate the analysis are available under an open license and can be found at https://github.com/alicelixuan/AMR-redispatch

\subsection*{Declaration of use of generative AI}
During the preparation of this work, the author used Claude to assist with writing parts of the code for the analysis. The author takes full responsibility for the content of the published article.

\subsection*{Acknowledgements}
The author thanks Clemens Stiewe, Jorge Sánchez Canales, and Lion Hirth for their valuable feedback. This work is supported by the German Federal Ministry of Research, Technology and Space (BMFTR) via the ARIADNE Project (FKZ 03SFK5K0-2).

\clearpage
\section*{Appendix}
\subsection*{A. Summary statistics}
{
\scriptsize
\setlength{\tabcolsep}{2.5pt}
\renewcommand{\arraystretch}{1.15}

\begin{longtable}{@{}p{2.7cm}p{1.65cm}rrrrrrr@{}}

\caption{Summary statistics by TSO control area}
\label{tab:summary_zonal}\\

\toprule
Variable & Control area & Min & 25\% & 50\% & Mean & 75\% & Max & Std. dev. \\
\midrule
\endfirsthead

\toprule
Variable & Control area & Min & 25\% & 50\% & Mean & 75\% & Max & Std. dev. \\
\midrule
\endhead

\midrule
\multicolumn{9}{r}{\textit{continued on next page}}\\
\endfoot

\bottomrule
\endlastfoot

\multirow{5}{2.7cm}{AMR usage}
 & 50Hertz & 0 & 0 & 175 & 742 & 998 & 16,290 & 1,292 \\
 & Amprion & 0 & 0 & 0 & 289 & 241 & 8,787 & 683 \\
 & TenneT DE & 0 & 0 & 165 & 645 & 877 & 9,551 & 1,055 \\
 & TransnetBW & 0 & 0 & 0 & 68 & 51 & 2,253 & 157 \\
 & \textit{All zones (pooled)} & 0 & 0 & 18 & 436 & 414 & 16,290 & 944 \\

\cmidrule(lr){1-9}

\multirow{5}{2.7cm}{Redispatch volume}
 & 50Hertz & 0 & 0 & 0 & 386 & 425 & 7,857 & 762 \\
 & Amprion & 0 & 0 & 28 & 603 & 873 & 6,976 & 962 \\
 & TenneT DE & 0 & 0 & 211 & 882 & 1,190 & 11,991 & 1,412 \\
 & TransnetBW & 0 & 0 & 54 & 365 & 546 & 3,905 & 571 \\
 & \textit{All zones (pooled)} & 0 & 0 & 49 & 559 & 703 & 11,991 & 1,000 \\

\cmidrule(lr){1-9}

\multirow{5}{2.7cm}{Redispatch volume, up}
 & 50Hertz & 0 & 0 & 0 & 19 & 0 & 2,063 & 105 \\
 & Amprion & 0 & 0 & 0 & 556 & 773 & 6,688 & 925 \\
 & TenneT DE & 0 & 0 & 0 & 142 & 156 & 4,064 & 296 \\
 & TransnetBW & 0 & 0 & 26 & 348 & 491 & 3,769 & 567 \\
 & \textit{All zones (pooled)} & 0 & 0 & 0 & 266 & 244 & 6,688 & 601 \\

\cmidrule(lr){1-9}

\multirow{5}{2.7cm}{Redispatch volume, down}
 & 50Hertz & 0 & 0 & 0 & 367 & 377 & 7,857 & 751 \\
 & Amprion & 0 & 0 & 0 & 46 & 0 & 2,423 & 173 \\
 & TenneT DE & 0 & 0 & 108 & 740 & 1,001 & 9,507 & 1,219 \\
 & TransnetBW & 0 & 0 & 0 & 16 & 0 & 656 & 55 \\
 & \textit{All zones (pooled)} & 0 & 0 & 0 & 292 & 125 & 9,507 & 779 \\

\cmidrule(lr){1-9}

\multirow{5}{2.7cm}{Grid load}
 & 50Hertz & 6,649 & 10,038 & 11,544 & 11,487 & 12,787 & 16,900 & 1,863 \\
 & Amprion & 6,633 & 16,227 & 18,920 & 18,886 & 21,409 & 27,550 & 3,227 \\
 & TenneT DE & 9,383 & 13,673 & 15,931 & 15,974 & 18,156 & 22,932 & 2,810 \\
 & TransnetBW & 3,496 & 5,349 & 6,436 & 6,518 & 7,634 & 10,400 & 1,384 \\
 & \textit{All zones (pooled)} & 3,496 & 8,695 & 13,103 & 13,216 & 17,269 & 27,550 & 5,275 \\

\cmidrule(lr){1-9}

\multirow{5}{2.7cm}{Offshore wind capacity}
 & 50Hertz & 1,093 & 1,352 & 1,828 & 1,561 & 1,828 & 1,828 & 300 \\
 & Amprion & 0 & 0 & 0 & 0 & 0 & 0 & 0 \\
 & TenneT DE & 7,040 & 7,110 & 7,387 & 7,248 & 7,391 & 7,398 & 155 \\
 & TransnetBW & 0 & 0 & 0 & 0 & 0 & 0 & 0 \\
 & \textit{All zones (pooled)} & 0 & 0 & 1,093 & 2,202 & 7,040 & 7,398 & 2,987 \\

\cmidrule(lr){1-9}

\multirow{5}{2.7cm}{Onshore wind capacity}
 & 50Hertz & 19,735 & 20,756 & 20,794 & 20,848 & 21,617 & 22,629 & 604 \\
 & Amprion & 10,874 & 11,620 & 12,138 & 11,962 & 12,737 & 13,939 & 630 \\
 & TenneT DE & 22,976 & 23,464 & 25,235 & 24,864 & 26,945 & 28,960 & 1,539 \\
 & TransnetBW & 1,674 & 1,674 & 1,749 & 1,761 & 1,893 & 2,165 & 87 \\
 & \textit{All zones (pooled)} & 1,674 & 10,874 & 19,735 & 14,859 & 22,976 & 28,960 & 8,931 \\

\cmidrule(lr){1-9}

\multirow{5}{2.7cm}{Solar capacity}
 & 50Hertz & 15,753 & 16,250 & 20,738 & 18,886 & 21,413 & 25,235 & 2,501 \\
 & Amprion & 12,895 & 14,475 & 17,563 & 16,914 & 20,946 & 24,487 & 2,983 \\
 & TenneT DE & 21,816 & 24,404 & 29,006 & 27,897 & 33,673 & 40,140 & 4,355 \\
 & TransnetBW & 7,280 & 7,937 & 9,294 & 8,901 & 10,376 & 13,398 & 1,153 \\
 & \textit{All zones (pooled)} & 7,280 & 12,895 & 17,563 & 18,150 & 21,816 & 40,140 & 7,383 \\

\cmidrule(lr){1-9}

\multicolumn{2}{l}{Solar irradiance (W/m²)}
 & 0.0 & 0.0 & 7.6 & 149.7 & 245.9 & 910.9 & 219.9 \\

\multicolumn{2}{l}{Wind speed (m/s)}
 & 1.4 & 4.7 & 6.1 & 6.4 & 7.6 & 18.9 & 2.3 \\

\end{longtable}

\noindent
\begin{minipage}{0.95\textwidth}
\scriptsize
\emph{Notes:} Statistics are computed per TSO control area over the sample period from June 9th 2022 to end of 2025. All variables are measured at hourly granularity. AMR usage is in MW, redispatch volume in MWh, load and capacity variables in MW. Solar irradiance is hourly mean incoming shortwave radiation at the Earth's surface, measured in W/m². Solar irradiance and hourly mean surface-layer wind speed (m/s) do not vary by control area and are therefore reported once.
\end{minipage}
}
{
\scriptsize
\setlength{\tabcolsep}{5pt}
\renewcommand{\arraystretch}{1.2}

\begin{longtable}{lrrrrrrr}
\caption{Summary statistics of hourly candidate control variables:\\neighbor-zone load (MW) and variable renewable generation (MWh)}
\small
\label{tab:summary_common}\\

\toprule
 & Min & 25\% & 50\% & Mean & 75\% & Max & Std. dev. \\
\midrule
\endfirsthead

\toprule
 & Min & 25\% & 50\% & Mean & 75\% & Max & Std. dev. \\
\midrule
\endhead

\midrule
\multicolumn{8}{r}{\textit{continued on next page}}\\
\endfoot

\bottomrule
\endlastfoot

Load AT & 4,014 & 5,686 & 6,692 & 6,730 & 7,613 & 10,444 & 1,292 \\
Load BE & 5,977 & 8,075 & 9,088 & 9,116 & 10,069 & 13,192 & 1,321 \\
Load CH & 2,743 & 6,258 & 6,999 & 6,982 & 7,696 & 10,437 & 1,063 \\
Load CZ & 3,828 & 6,002 & 6,999 & 7,014 & 7,941 & 10,861 & 1,305 \\
Load DK1 & 1,318 & 2,179 & 2,528 & 2,526 & 2,857 & 3,888 & 454 \\
Load DK2 & 914 & 1,355 & 1,575 & 1,580 & 1,792 & 5,968 & 303 \\
Load FR & 28,744 & 41,514 & 47,189 & 48,781 & 54,644 & 86,573 & 10,031 \\
Load HR & 1,128 & 1,750 & 2,062 & 2,065 & 2,348 & 3,340 & 397 \\
Load HU & 2,216 & 4,268 & 4,822 & 4,879 & 5,424 & 7,450 & 811 \\
Load NL & 5,440 & 11,158 & 12,610 & 12,728 & 13,914 & 19,463 & 2,049 \\
Load NO1 & 1,693 & 2,795 & 3,559 & 3,846 & 4,843 & 8,166 & 1,275 \\
Load NO2 & 1,595 & 3,472 & 3,890 & 4,026 & 4,551 & 6,434 & 691 \\
Load NO3 & 2,010 & 2,805 & 3,149 & 3,212 & 3,602 & 4,721 & 496 \\
Load NO4 & 1,430 & 1,943 & 2,292 & 2,298 & 2,636 & 3,407 & 399 \\
Load NO5 & 847 & 1,647 & 1,866 & 1,902 & 2,154 & 4,326 & 340 \\
Load PL & 10,582 & 16,120 & 18,739 & 18,691 & 21,079 & 28,304 & 3,191 \\
Load RO & 0 & 5,389 & 6,066 & 6,137 & 6,838 & 8,907 & 967 \\
Load SI & 93 & 1,200 & 1,406 & 1,423 & 1,630 & 2,316 & 298 \\
Load SK & 1,899 & 2,580 & 2,918 & 2,924 & 3,231 & 4,170 & 450 \\

Offshore wind DK1 & 0 & 218 & 600 & 655 & 1,044 & 1,925 & 474 \\
Offshore wind DK2 & 0 & 94 & 316 & 389 & 676 & 976 & 316 \\
Offshore wind NL & 0 & 423 & 1,196 & 1,493 & 2,365 & 4,523 & 1,215 \\
Offshore wind NO2 & 0 & 0 & 0 & 0 & 0 & 5 & 1 \\

Onshore wind AT & 4 & 206 & 638 & 948 & 1,492 & 3,645 & 894 \\
Onshore wind BE & 1 & 168 & 437 & 624 & 918 & 2,724 & 579 \\
Onshore wind CH & 0 & 6 & 12 & 17 & 25 & 76 & 15 \\
Onshore wind CZ & 1 & 30 & 59 & 74 & 105 & 298 & 57 \\
Onshore wind DK1 & 0 & 306 & 775 & 1,013 & 1,523 & 3,540 & 855 \\
Onshore wind DK2 & 3 & 59 & 133 & 176 & 268 & 605 & 141 \\
Onshore wind HR & 0 & 81 & 222 & 304 & 498 & 1,117 & 259 \\
Onshore wind HU & 0 & 10 & 42 & 67 & 105 & 304 & 69 \\
Onshore wind NL & 6 & 281 & 648 & 846 & 1,313 & 2,545 & 666 \\
Onshore wind NO1 & 0 & 29 & 84 & 113 & 180 & 380 & 98 \\
Onshore wind NO3 & 1 & 227 & 513 & 642 & 1,022 & 1,997 & 485 \\
Onshore wind NO4 & 0 & 161 & 310 & 352 & 518 & 1,039 & 230 \\
Onshore wind PL & 20 & 911 & 1,916 & 2,459 & 3,574 & 8,897 & 1,944 \\
Onshore wind RO & 0 & 219 & 541 & 733 & 1,087 & 2,762 & 639 \\
Onshore wind SI & 0 & 0 & 0 & 1 & 1 & 5 & 1 \\


Solar AT & 0 & 0 & 23 & 463 & 554 & 4,672 & 826 \\
Solar BE & 0 & 0 & 20 & 945 & 1,386 & 7,726 & 1,523 \\
Solar CH & 0 & 0 & 4 & 471 & 642 & 4,018 & 793 \\
Solar CZ & 0 & 0 & 39 & 405 & 592 & 3,231 & 647 \\
Solar DK1 & 0 & 0 & 14 & 289 & 395 & 2,752 & 479 \\
Solar DK2 & 0 & 0 & 5 & 113 & 158 & 908 & 185 \\
Solar FR & 0 & 0 & 367 & 2,759 & 4,991 & 19,434 & 3,832 \\
Solar HR & 0 & 0 & 2 & 31 & 37 & 361 & 57 \\
Solar HU & 0 & 0 & 21 & 599 & 989 & 4,045 & 908 \\
Solar NL & 0 & 0 & 2 & 58 & 88 & 428 & 92 \\
Solar NO1 & 0 & 0 & 0 & 0 & 0 & 7 & 1 \\
Solar PL & 0 & 0 & 140 & 1,780 & 2,860 & 13,682 & 2,685 \\
Solar RO & 0 & 0 & 6 & 226 & 384 & 1,830 & 346 \\
Solar SI & 0 & 0 & 3 & 57 & 78 & 910 & 101 \\
Solar SK & 0 & 0 & 2 & 60 & 96 & 383 & 93 \\


\end{longtable}

\noindent
\begin{minipage}{0.9\textwidth}
\scriptsize
\emph{Notes:} Suffixes are bidding-zone codes.
\end{minipage}
}

\subsection*{B. Complete regression results and additional robustness}
\begin{table}[htbp]
    \centering
    \caption{Effect of AMR on aggregate German redispatch (main specification)}
    \label{tab:results_main}

    \scriptsize
    \setlength{\tabcolsep}{4pt}
    \renewcommand{\arraystretch}{1.1}

    \begin{tabular}{lc}
        \toprule
        & (1) PDS (1-SE) adjusted \\
        \midrule
        AMR
        & $0.199^{***}$ \\
        & $(0.034)$ \\[0.2em]

        Grid load
        & $0.048^{***}$ \\
        & $(0.009)$ \\[0.2em]

        Onshore wind potential
        & $-0.0038^{***}$ \\
        & $(0.0010)$ \\[0.2em]

        Offshore wind potential
        & $0.030^{***}$ \\
        & $(0.003)$ \\[0.2em]

        Onshore wind generation, PL
        & $0.052^{***}$ \\
        & $(0.013)$ \\[0.2em]

        Onshore wind generation, DK1
        & $0.148^{***}$ \\
        & $(0.033)$ \\[0.2em]

        Onshore wind generation, DK2
        & $1.941^{***}$ \\
        & $(0.247)$ \\[0.2em]

        \midrule

        TSO fixed effects
        & Yes \\

        Hour-of-day fixed effects
        & Yes \\

        Observations
        & $123{,}689$ \\

        Within $R^2$
        & $0.408$ \\

        \bottomrule
    \end{tabular}

    \vspace{0.4em}
    \begin{minipage}{0.90\textwidth}
        \scriptsize
        \textit{Notes:} The dependent variable is hourly control-area aggregate redispatch volume in MWh. Hour-of-day coefficients and the intercept are omitted for brevity. Driscoll-Kraay standard errors are reported in parentheses. $^{*}\,p<0.05$, $^{**}\,p<0.01$, $^{***}\,p<0.001$.
    \end{minipage}
\end{table}
\begin{table}[htbp]
    \centering
    \caption{Effect of AMR on upward, downward, and total German redispatch}
    \label{tab:redispatch_direction}

    \scriptsize
    \setlength{\tabcolsep}{4pt}
    \renewcommand{\arraystretch}{1.1}

    \begin{tabular}{lccc}
        \toprule
        & (1) Upward
        & (2) Downward
        & (3) Total \\
        \midrule

        AMR
        & $0.012$
        & $0.187^{***}$
        & $0.199^{***}$ \\
        & $(0.016)$
        & $(0.026)$
        & $(0.034)$ \\[0.2em]

        Grid load
        & $0.025^{***}$
        & $0.023^{***}$
        & $0.048^{***}$ \\
        & $(0.005)$
        & $(0.005)$
        & $(0.009)$ \\[0.2em]

        Onshore wind potential
        & $-0.000$
        & $-0.004^{***}$
        & $-0.004^{***}$ \\
        & $(0.001)$
        & $(0.001)$
        & $(0.001)$ \\[0.2em]

        Offshore wind potential
        & $-0.007^{**}$
        & $0.037^{***}$
        & $0.030^{***}$ \\
        & $(0.002)$
        & $(0.003)$
        & $(0.003)$ \\[0.2em]

        Wind onshore PL
        & $0.040^{***}$
        & $0.013^{*}$
        & $0.052^{***}$ \\
        & $(0.007)$
        & $(0.006)$
        & $(0.013)$ \\[0.2em]

        Wind onshore DK\_1
        & $0.101^{***}$
        & $0.047^{*}$
        & $0.148^{***}$ \\
        & $(0.016)$
        & $(0.020)$
        & $(0.033)$ \\[0.2em]

        Wind onshore DK\_2
        & $1.059^{***}$
        & $0.883^{***}$
        & $1.941^{***}$ \\
        & $(0.121)$
        & $(0.135)$
        & $(0.247)$ \\

        \midrule

        AMR effect per IQR (414 MW)
        & $4.8$ MW
        & $77.5$ MW
        & $82.3$ MW \\

        TSO fixed effects
        & Yes & Yes & Yes \\

        Hour-of-day fixed effects
        & Yes & Yes & Yes \\

        Observations
        & $123{,}689$
        & $123{,}689$
        & $123{,}689$ \\

        Within $R^2$
        & $0.2344$
        & $0.3366$
        & $0.4079$ \\

        \bottomrule
    \end{tabular}

    \vspace{0.4em}
    \begin{minipage}{0.94\textwidth}
        \scriptsize
        \textit{Notes:} The dependent variables are hourly upward redispatch volume in column (1), downward redispatch volume in column (2), and total redispatch volume in column (3), all measured in MWh. The AMR effect per IQR reports the implied change in redispatch associated with a 414 MW increase in AMR. Driscoll-Kraay standard errors are reported in parentheses. Hour-of-day coefficients and the intercept are omitted for brevity. $^{*}\,p<0.05$, $^{**}\,p<0.01$, $^{***}\,p<0.001$.
    \end{minipage}
\end{table}

\FloatBarrier
\subsection*{C. Additional figures}
\begin{center}

\includegraphics[
    width=1\textwidth,
    height=0.45\textheight,
    keepaspectratio
]{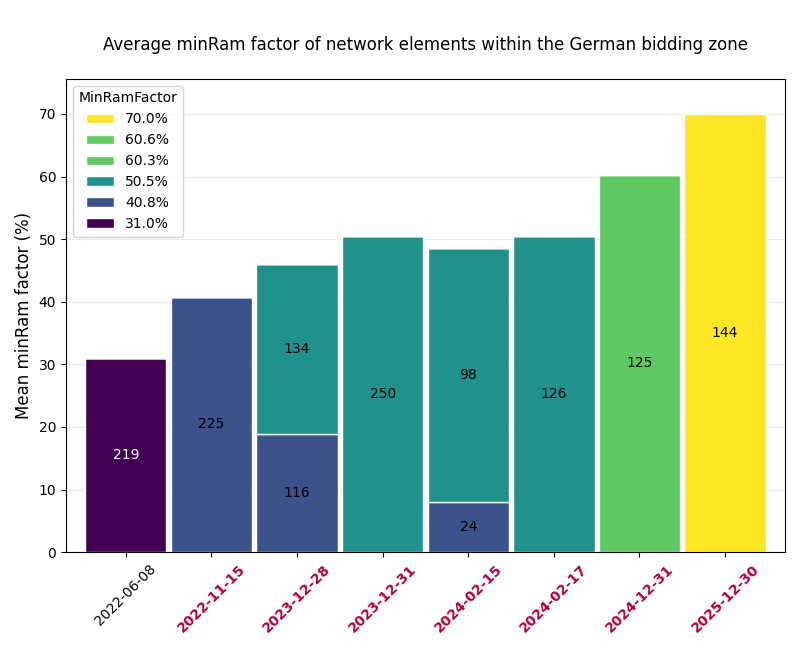}

\captionof{figure}{Progression of minRAM factors applied to base-case critical network elements (CNEs) within the German bidding zone over the implementation period: the height of each bar shows the average minRAM factor for all CNEs at selected time points (in red) when minRAM factor changes for more than 10 CNEs. The start of the sample period is also shown. The numbers in each bar denote the number of CNEs in the data.}
\label{fig:av_minRAM}
\end{center}

\begin{center}

\includegraphics[
    width=0.97\textwidth,
    height=0.42\textheight,
    keepaspectratio
]{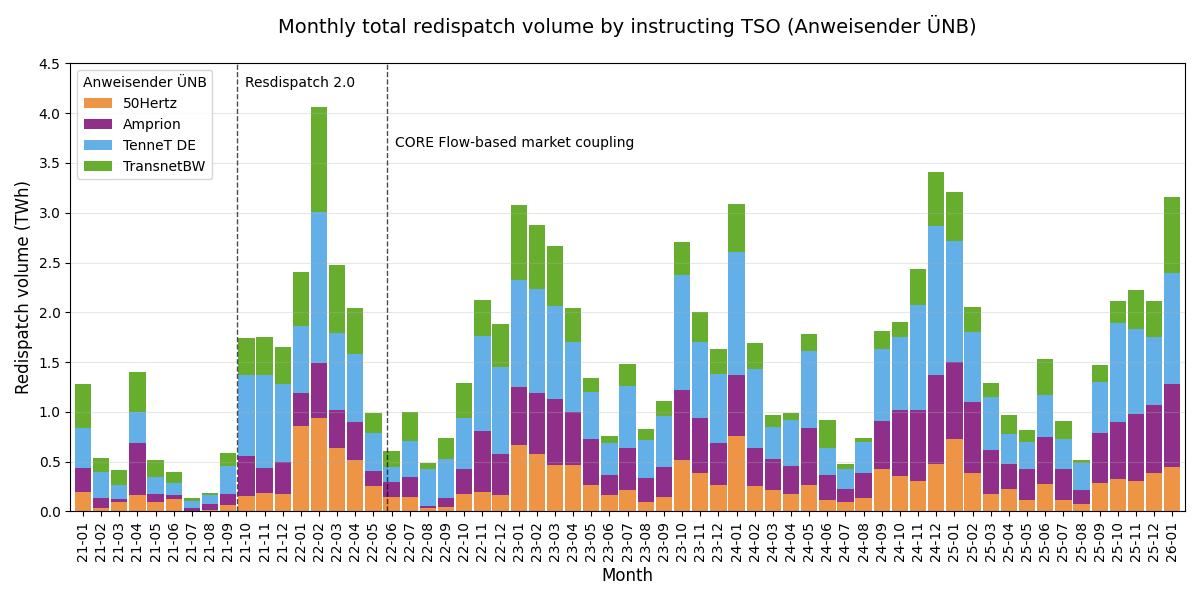}

\captionof{figure}{Total redispatch volume by instructing TSO. Data transformed from redispatch orders \citep{NetztransparenzSystemdienstleistungenBetriebsfuehrung} to hourly time series and aggregated by month, including both up- and downward redispatch.}
\label{fig:redispatch}

\vspace{0.3cm}

\includegraphics[
    width=0.97\textwidth,
    height=0.42\textheight,
    keepaspectratio
]{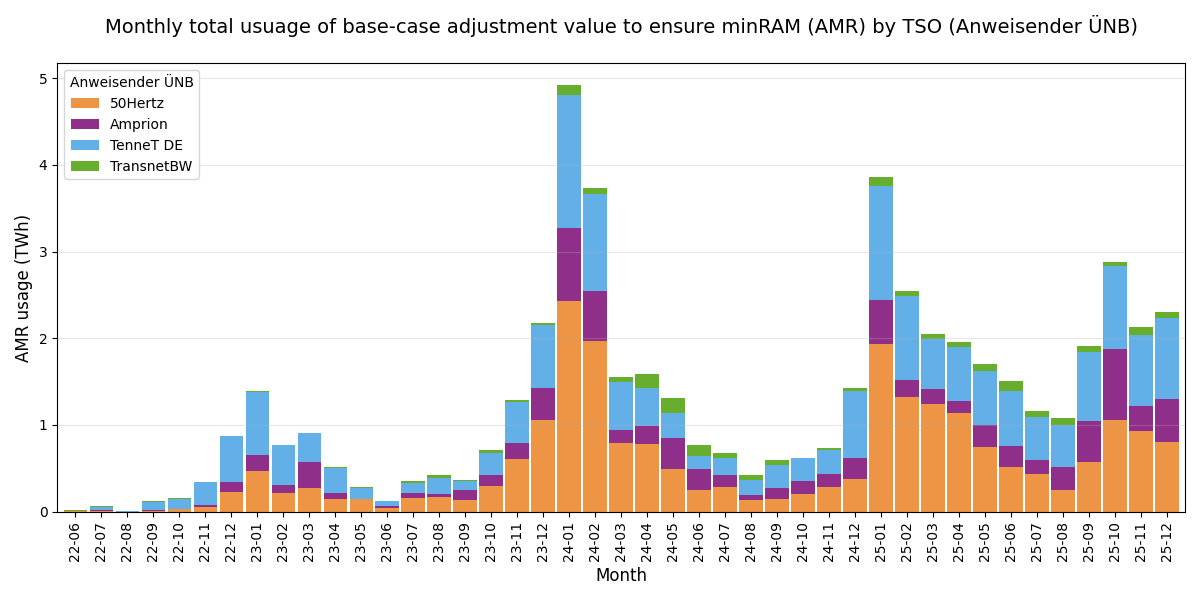}

\captionof{figure}{Total base-case AMR usage by TSO. Data obtained from hourly flow-based input data \citep{jointallocationofficeJointAllocationOffice2025} and aggregated by month, for all CNEs of each hour.}
\label{fig:amr}

\end{center}

\begin{figure}
    \centering
    \includegraphics[width=1\textwidth,trim={0cm 0cm 0cm 0cm},clip]{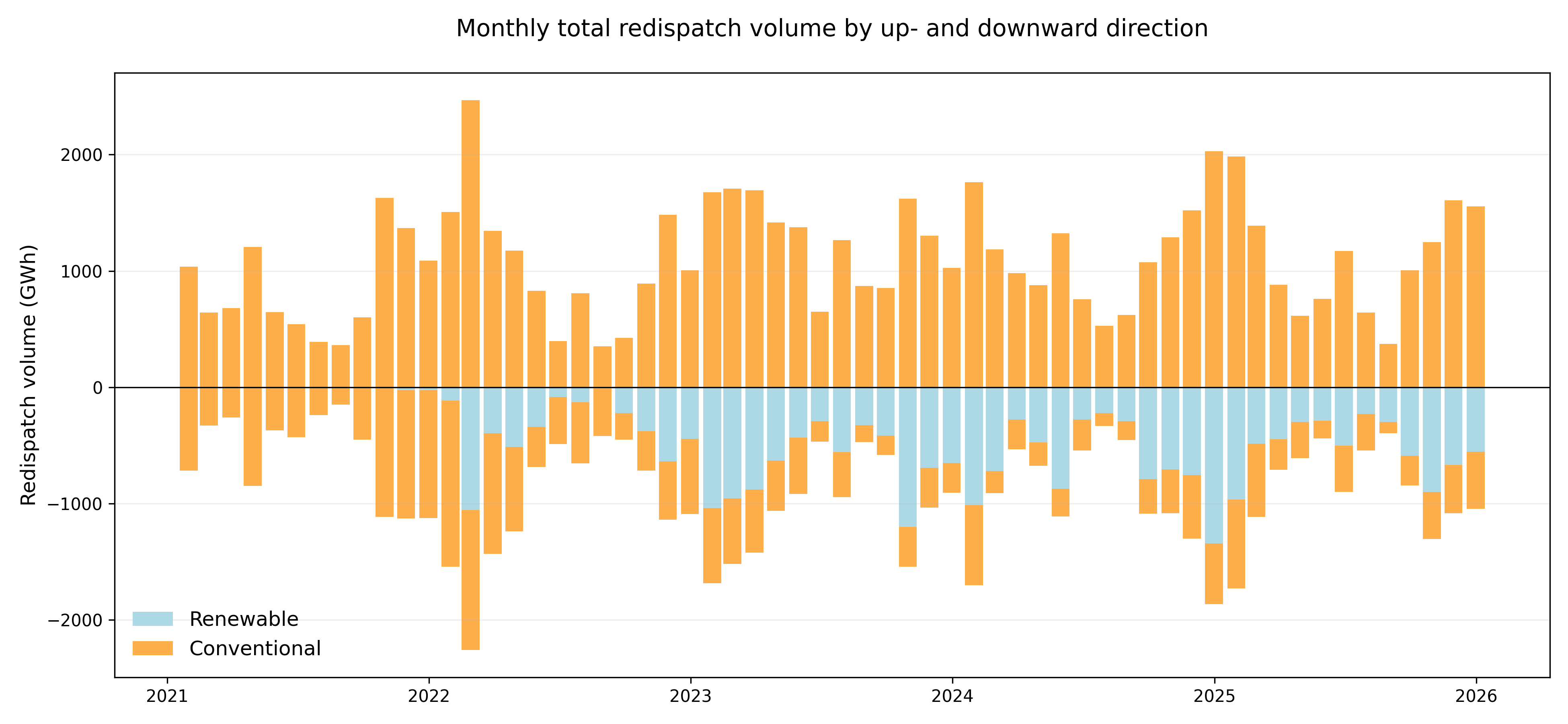}
    \caption{Monthly total redispatch volume by up- and downward direction and by conventional vs. renewable technology type. Individual generation units in redispatch order data \citep{NetztransparenzSystemdienstleistungenBetriebsfuehrung} are mapped to either technology type.} 
    \label{fig:rdp_res}
\end{figure}
\FloatBarrier
\bibliography{Redispatch}

@misc{20250117CECCM,
  title = {{{TSO}} Formal Submission on Day-Ahead Capacity Calculation Methodology of the {{Central Europe}} Capacity Calculation Region},
  author = {{ENTSO-E}},
  year = 2025,
  urldate = {2026-08-10},
  howpublished = {\url{https://eepublicdownloads.entsoe.eu/clean-documents/nc-tasks/20250117\%20CE\%20DA\%20CCM\%20TSO\%20formal\%20submission\%20pdf\_.pdf}}
}

@article{belloniHighDimensionalMethodsInference2014,
  title = {High-{{Dimensional Methods}} and {{Inference}} on {{Structural}} and {{Treatment Effects}}},
  author = {Belloni, Alexandre and Chernozhukov, Victor and Hansen, Christian},
  year = 2014,
  month = may,
  journal = {Journal of Economic Perspectives},
  volume = {28},
  number = {2},
  pages = {29--50},
  issn = {0895-3309},
  doi = {10.1257/jep.28.2.29},
  urldate = {2026-08-25},
  langid = {english}
}

@article{bucksteegImpactsRenewablesGeneration2015,
  title = {Impacts of Renewables Generation and Demand Patterns on Net Transfer Capacity: Implications for Effectiveness of Market Splitting in {{Germany}}},
  shorttitle = {Impacts of Renewables Generation and Demand Patterns on Net Transfer Capacity},
  author = {Bucksteeg, Michael and Trepper, Katrin and Weber, Christoph},
  year = 2015,
  journal = {IET Generation, Transmission \& Distribution},
  volume = {9},
  number = {12},
  pages = {1510--1518},
  issn = {1751-8695},
  doi = {10.1049/iet-gtd.2014.1063},
  urldate = {2026-09-03},
  langid = {english}
}

@article{bucksteegImprovingFlowbasedMarket2024,
  title = {Improving Flow-Based Market Coupling by Integrating Redispatch potential\texthorizontalbar{{Evidence}} from a Large-Scale Model},
  author = {Bucksteeg, Michael and Voswinkel, Simon and Blumberg, Gerald},
  year = 2024,
  month = may,
  journal = {Energy Policy},
  volume = {188},
  pages = {114093},
  issn = {0301-4215},
  doi = {10.1016/j.enpol.2024.114093},
  urldate = {2026-08-13}
}

@misc{BundesnetzagenturBeschlusskammer6Festlegungsverfahren,
  title = {Festlegungsverfahren Zur {{Netzbetreiberkoordinierung}} Bei Der {{Durchf\"uhrung}} von {{Redispatch-Ma\ss nahmen}}},
  author = {{Bundesnetzagentur}},
  year = 2021,
  urldate = {2026-08-24},
  howpublished = {\url{https://www.bundesnetzagentur.de/DE/Beschlusskammern/1\_GZ/BK6-GZ/2020/BK6-20-060/BK6-20-060\_beschluss.html}}
}

@misc{BundesnetzagenturNetzengpassmanagement,
  title = {Bundesnetzagentur - {{Netzengpassmanagement}}},
  author = {{Bundesnetzagentur}},
  year = 2026,
  urldate = {2026-09-07},
  howpublished = {\url{https://www.bundesnetzagentur.de/DE/Fachthemen/ElektrizitaetundGas/Versorgungssicherheit/Netzengpassmanagement/start.html}}
}

@misc{CommissionRegulationEU2015,
  title = {Commission {{Regulation}} ({{EU}}) 2015/1222 of 24 {{July}} 2015 Establishing a Guideline on Capacity Allocation and Congestion Management ({{Text}} with {{EEA}} Relevance)},
  author = {{EU}},
  year = 2015,
  urldate = {2026-08-16},
  howpublished = {\url{http://data.europa.eu/eli/reg/2015/1222/oj}},
  langid = {english}
}

@article{egererTwoPriceZones2016,
  title = {Two Price Zones for the {{German}} Electricity Market --- {{Market}} Implications and Distributional Effects},
  author = {Egerer, Jonas and Weibezahn, Jens and Hermann, Hauke},
  year = 2016,
  month = sep,
  journal = {Energy Economics},
  volume = {59},
  pages = {365--381},
  issn = {0140-9883},
  doi = {10.1016/j.eneco.2016.08.002},
  urldate = {2026-08-13}
}

@misc{entso-eENTSOETransparencyPlatform2025,
  title = {{{ENTSO-E Transparency Platform}}},
  author = {{ENTSO-E}},
  year = 2026,
  howpublished = {\url{https://transparency.entsoe.eu/}}
}

@article{fellingAssessingImprovedPrice2023,
  title = {Assessing {{Improved Price Zones}} in {{Europe}}: {{Flow-Based Market Coupling}} in {{Central Western Europe}} in {{Focus}}},
  shorttitle = {Assessing {{Improved Price Zones}} in {{Europe}}},
  author = {Felling, Tim and Felten, Bj{\"o}rn and Osinski, Paul and Weber, Christoph},
  year = 2023,
  month = nov,
  journal = {The Energy Journal},
  volume = {44},
  number = {6},
  pages = {71--112},
  publisher = {SAGE Publications},
  issn = {0195-6574},
  doi = {10.5547/01956574.44.6.tfel},
  urldate = {2026-07-09},
  langid = {english}
}

@article{feltenFlowbasedMarketCoupling2021,
  title = {The Flow-Based Market Coupling Domain - {{Why}} We Can't Get It Right},
  author = {Felten, Bj{\"o}rn and Osinski, Paul and Felling, Tim and Weber, Christoph},
  year = 2021,
  month = jun,
  journal = {Utilities Policy},
  volume = {70},
  pages = {101136},
  issn = {0957-1787},
  doi = {10.1016/j.jup.2020.101136},
  urldate = {2026-07-03}
}

@article{finckImpactFlowBased2021,
  title = {Impact of {{Flow Based Market Coupling}} on the {{European Electricity Markets}}},
  author = {Finck, Rafael},
  year = 2021,
  month = jun,
  journal = {Sustainability Management Forum \textbar{} NachhaltigkeitsManagementForum},
  volume = {29},
  number = {2},
  pages = {173--186},
  issn = {2522-5995},
  doi = {10.1007/s00550-021-00520-w},
  urldate = {2026-08-27},
  langid = {english}
}

@article{henneauxImpactAssessmentMinimum2021a,
  title = {Impact Assessment of a Minimum Threshold on Cross-Zonal Capacity in a Flow-Based Market},
  author = {Henneaux, Pierre and Lamprinakos, Panagiotis and {de Maere d'Aertrycke}, Gauthier and Karoui, Karim},
  year = 2021,
  month = jan,
  journal = {Electric Power Systems Research},
  volume = {190},
  pages = {106693},
  issn = {0378-7796},
  doi = {10.1016/j.epsr.2020.106693},
  urldate = {2026-09-03}
}

@article{hirthHowAggregateElectricity2024,
  title = {How Aggregate Electricity Demand Responds to Hourly Wholesale Price Fluctuations},
  author = {Hirth, Lion and Khanna, Tarun M. and Ruhnau, Oliver},
  year = 2024,
  journal = {Energy Economics},
  volume = {135},
  pages = {107652},
  issn = {0140-9883},
  doi = {10.1016/j.eneco.2024.107652},
  urldate = {2025-05-21}
}

@techreport{hirthMarketBasedRedispatchZonal2019,
  type = {Working {{Paper}}},
  title = {Market-{{Based Redispatch}} in {{Zonal Electricity Markets}}: {{Inc-Dec Gaming}} as a {{Consequence}} of {{Inconsistent Power Market Design}} (Not {{Market Power}})},
  shorttitle = {Market-{{Based Redispatch}} in {{Zonal Electricity Markets}}},
  author = {Hirth, Lion and Schlecht, Ingmar},
  year = 2019,
  institution = {Kiel, Hamburg: ZBW -- Leibniz Information Centre for Economics},
  urldate = {2026-02-19},
  copyright = {http://www.econstor.eu/dspace/Nutzungsbedingungen},
  langid = {english}
}

@article{holmbergComparisonCongestionManagement2015,
  title = {Comparison of {{Congestion Management Techniques}}: {{Nodal}}, {{Zonal}} and {{Discriminatory Pricing}}},
  shorttitle = {Comparison of {{Congestion Management Techniques}}},
  author = {Holmberg, P{\"a}r and Lazarczyk, Ewa},
  year = 2015,
  journal = {The Energy Journal},
  volume = {36},
  number = {2},
  eprint = {24695751},
  eprinttype = {jstor},
  pages = {145--166},
  publisher = {International Association for Energy Economics},
  issn = {0195-6574},
  urldate = {2026-08-20}
}

@article{InsightsGermanysFuture,
  title = {Insights on {{Germany}}'s {{Future Congestion Management}} from a {{Multi-Model Approach}}},
  author = {Hladik, Dirk and Fraunholz, Christoph and K{\"u}hnbach, Matthias and Manz, Pia and Kunze, Robert},
  year = 2020,
  journal = {Energies},
  volume = {13},
  number = {16},
  issn = {4176},
  doi = {10.3390/en13164176},
  urldate = {2026-08-17}
}

@misc{jointallocationofficeJointAllocationOffice2025,
  title = {Joint {{Allocation Office Publication Tool}}},
  author = {{JAO}},
  year = 2026,
  howpublished = {https://publicationtool.jao.eu/}
}

@article{kenisRegulatoryIncentivesTransmission2024a,
  title = {Regulatory Incentives for Transmission System Operators under Flow-Based Market Coupling},
  author = {Kenis, Michiel and Bruninx, Kenneth and Delarue, Erik},
  year = 2024,
  month = apr,
  journal = {Utilities Policy},
  volume = {87},
  pages = {101707},
  issn = {0957-1787},
  doi = {10.1016/j.jup.2023.101707},
  urldate = {2026-08-03}
}

@misc{knorrZonalVsNodal2024,
  title = {Zonal vs. {{Nodal Pricing}}: {{An Analysis}} of {{Different Pricing Rules}} in the {{German Day-Ahead Market}}},
  shorttitle = {Zonal vs. {{Nodal Pricing}}},
  author = {Kn{\"o}rr, Johannes and Bichler, Martin and Dobos, Teodora},
  year = 2024,
  publisher = {arXiv},
  doi = {10.48550/ARXIV.2403.09265},
  urldate = {2026-07-03},
  copyright = {Creative Commons Attribution 4.0 International},
  langid = {english}
}

@inproceedings{marjanovicImpactEnlargementFlow2018,
  title = {Impact of an {{Enlargement}} of the {{Flow Based Region}} in {{Continental Europe}}},
  booktitle = {2018 15th {{International Conference}} on the {{European Energy Market}} ({{EEM}})},
  author = {Marjanovic, I. and vom Stein, D. and {van Bracht}, N. and Moser, A.},
  year = 2018,
  month = jun,
  pages = {1--5},
  issn = {2165-4093},
  doi = {10.1109/EEM.2018.8470008},
  urldate = {2026-08-17}
}

@inproceedings{matthesImpactMinimumRemaining2019,
  title = {Impact of a {{Minimum Remaining Available Margin Adjustment}} in {{Flow-Based Market Coupling}}},
  booktitle = {2019 {{IEEE Milan PowerTech}}},
  author = {Matthes, Bj{\"o}rn and Spieker, Christopher and Klein, Dennis and Rehtanz, Christian},
  year = 2019,
  month = jun,
  pages = {1--6},
  doi = {10.1109/PTC.2019.8810504},
  urldate = {2026-08-17}
}

@misc{MERRA2Tavg1_2d_flx_Nx2d1HourlyTimeAveragedSingleLevelAssimilationSurface,
  title = {{{MERRA-2}} tavg1\_2d\_flx\_{{Nx}}: 2d,1-{{Hourly}},{{Time-Averaged}},{{Single-Level}},{{Assimilation}},{{Surface Flux Diagnostics}} 0.625 x 0.5 Degree {{V5}}.12.4 ({{M2T1NXFLX}}) - {{NASA Open Data Portal}}},
  shorttitle = {{{MERRA-2}} tavg1\_2d\_flx\_{{Nx}}},
  author = {{NASA}},
  year = 2026,
  doi = {10.5067/7MCPBJ41Y0K6},
  urldate = {2026-08-16},
  howpublished = {\url{https://data.nasa.gov/dataset/merra-2-tavg1-2d-flx-nx-2d1-hourlytime-averagedsingle-levelassimilationsurface-flux-diagno-f916f}},
  langid = {english}
}

@misc{MonitoringReport2025,
  title = {Monitoring {{Report}} 2025 in Accordance with Section 63(3) in Conjunction with Section 35 of the {{Energy Industry Act}} ({{EnWG}}) and Section 48(3) in Conjunction with Section 53(3) of the {{Competition Act}} ({{GWB}})},
  author = {{Bundesnetzagentur} and {Bundeskartellamt}},
  year = 2026,
  howpublished = {\url{https://data.bundesnetzagentur.de/Bundesnetzagentur/SharedDocs/Downloads/EN/Areas/ElectricityGas/CollectionCompanySpecificData/Monitoring/MonitoringReport2025.pdf}},
  langid = {english}
}

@misc{NetztransparenzSystemdienstleistungenBetriebsfuehrung,
  title = {Redispatch},
  author = {{Netztransparenz}},
  year = 2026,
  urldate = {2026-08-16},
  howpublished = {\url{https://www.netztransparenz.de/de-de/Systemdienstleistungen/Betriebsfuehrung/Redispatch}}
}

@article{ovaereEffectFlowbasedMarket2023,
  title = {The Effect of Flow-Based Market Coupling on Cross-Border Exchange Volumes and Price Convergence in {{Central Western European}} Electricity Markets},
  author = {Ovaere, Marten and Kenis, Michiel and {Van den Bergh}, Kenneth and Bruninx, Kenneth and Delarue, Erik},
  year = 2023,
  month = feb,
  journal = {Energy Economics},
  volume = {118},
  pages = {106519},
  issn = {0140-9883},
  doi = {10.1016/j.eneco.2023.106519},
  urldate = {2026-07-03}
}

@article{poplavskayaIntegrationDayaheadMarket2020,
  title = {Integration of Day-Ahead Market and Redispatch to Increase Cross-Border Exchanges in the {{European}} Electricity Market},
  author = {Poplavskaya, Ksenia and Totschnig, Gerhard and Leimgruber, Fabian and Doorman, Gerard and Etienne, Gilles and {de Vries}, Laurens},
  year = 2020,
  month = nov,
  journal = {Applied Energy},
  volume = {278},
  pages = {115669},
  issn = {0306-2619},
  doi = {10.1016/j.apenergy.2020.115669},
  urldate = {2026-08-12}
}

@misc{RegulationEU20192019,
  title = {Regulation ({{EU}}) 2019/943 of the {{European Parliament}} and of the {{Council}} of 5 {{June}} 2019 on the Internal Market for Electricity (Recast) ({{Text}} with {{EEA}} Relevance.)},
  author = {{EU}},
  year = 2019,
  urldate = {2026-07-22},
  howpublished = {\url{http://data.europa.eu/eli/reg/2019/943/oj}},
  langid = {english}
}

@article{rubinRandomizationAnalysisExperimental1980,
  title = {Randomization {{Analysis}} of {{Experimental Data}}: {{The Fisher Randomization Test Comment}}},
  shorttitle = {Randomization {{Analysis}} of {{Experimental Data}}},
  author = {Rubin, Donald B.},
  year = 1980,
  journal = {Journal of the American Statistical Association},
  volume = {75},
  number = {371},
  eprint = {2287653},
  eprinttype = {jstor},
  pages = {591--593},
  publisher = {[American Statistical Association, Taylor \& Francis, Ltd.]},
  issn = {0162-1459},
  doi = {10.2307/2287653},
  urldate = {2026-09-07}
}

@techreport{schonheitELMODDocumentationModeling2020,
  type = {Working {{Paper}}},
  title = {{{ELMOD}} Documentation: {{Modeling}} of Flow-Based Market Coupling and Congestion Management},
  shorttitle = {{{ELMOD}} Documentation},
  author = {Sch{\"o}nheit, David and Hladik, Dirk and Hobbie, Hannes and M{\"o}st, Dominik},
  year = 2020,
  institution = {Working paper of the Chair of Energy Economics (TU Dresden)},
  urldate = {2026-07-29},
  copyright = {http://www.econstor.eu/dspace/Nutzungsbedingungen},
  langid = {english}
}

@article{schonheitFundamentalUnderstandingFlowbased2021,
  title = {Toward a Fundamental Understanding of Flow-Based Market Coupling for Cross-Border Electricity Trading},
  author = {Sch{\"o}nheit, David and Kenis, Michiel and Lorenz, Lisa and M{\"o}st, Dominik and Delarue, Erik and Bruninx, Kenneth},
  year = 2021,
  month = may,
  journal = {Advances in Applied Energy},
  volume = {2},
  pages = {100027},
  issn = {2666-7924},
  doi = {10.1016/j.adapen.2021.100027},
  urldate = {2026-07-09}
}

@article{schonheitImpactDifferentStrategies2020,
  title = {The Impact of Different Strategies for Generation Shift Keys ({{GSKs}}) on the Flow-Based Market Coupling Domain: {{A}} Model-Based Analysis of {{Central Western Europe}}},
  shorttitle = {The Impact of Different Strategies for Generation Shift Keys ({{GSKs}}) on the Flow-Based Market Coupling Domain},
  author = {Sch{\"o}nheit, David and Weinhold, Richard and Dierstein, Constantin},
  year = 2020,
  month = jan,
  journal = {Applied Energy},
  volume = {258},
  pages = {114067},
  issn = {0306-2619},
  doi = {10.1016/j.apenergy.2019.114067},
  urldate = {2026-07-29}
}

@article{schonheitMinimumTradingCapacities2021,
  title = {Do Minimum Trading Capacities for the Cross-Zonal Exchange of Electricity Lead to Welfare Losses?},
  author = {Sch{\"o}nheit, David and Dierstein, Constantin and M{\"o}st, Dominik},
  year = 2021,
  month = feb,
  journal = {Energy Policy},
  volume = {149},
  pages = {112030},
  issn = {0301-4215},
  doi = {10.1016/j.enpol.2020.112030},
  urldate = {2026-07-29}
}

@misc{SMARDDownloadMarket,
  title = {{{SMARD}} \textbar{} {{Download}} Market Data},
  author = {{Bundesnetzagentur}},
  year = 2026,
  urldate = {2026-08-16},
  howpublished = {\url{https://www.smard.de/en/downloadcenter/download-market-data/}}
}

@misc{SMARDNetzengpassmanagement,
  title = {{{SMARD}} \textbar{} {{Netzengpassmanagement}}},
  author = {{Bundesnetzagentur}},
  year = 2026,
  urldate = {2026-09-09},
  howpublished = {\url{https://www.smard.de/page/home/topic-article/46/213328/netzengpassmanagement}}
}

@misc{SMARDRedispatchKostena,
  title = {{{SMARD}} \textbar{} {{Redispatch}} - {{Kosten}} pro {{MWh}}},
  author = {{Bundesnetzagentur}},
  year = 2026,
  urldate = {2026-09-12},
  howpublished = {\url{https://www.smard.de/page/home/topic-article/211972/215340/redispatch-kosten-pro-mwh}}
}

@article{titzIdentifyingDriversMitigators2024,
  title = {Identifying Drivers and Mitigators for Congestion and Redispatch in the {{German}} Electric Power System with Explainable {{AI}}},
  author = {Titz, Maurizio and P{\"u}tz, Sebastian and Witthaut, Dirk},
  year = 2024,
  month = feb,
  journal = {Applied Energy},
  volume = {356},
  pages = {122351},
  issn = {0306-2619},
  doi = {10.1016/j.apenergy.2023.122351},
  urldate = {2026-08-13}
}

@inproceedings{weberImpactRegionalTimeDependent2023,
  title = {Impact of a {{Regional-}} and {{Time-Dependent Minimum Remaining Available Margin}} on {{Market Coupling}} and {{Redispatch}}},
  booktitle = {2023 {{International Conference}} on {{Future Energy Solutions}} ({{FES}})},
  author = {Weber, Lukas and Meuthrath, David and Pacco, Kevin and Moser, Albert},
  year = 2023,
  month = jun,
  pages = {1--6},
  doi = {10.1109/FES57669.2023.10183037},
  urldate = {2026-08-10}
}

\end{document}